\documentclass{article}
\usepackage{graphicx} 
\usepackage{amsmath, amssymb} 
\usepackage{float} 
\usepackage{hyperref} 
\usepackage{geometry} 
\usepackage{caption} 
\usepackage{booktabs} 
\usepackage{adjustbox} 
\usepackage{subfigure}
\usepackage{graphicx}
\usepackage{subcaption} 
\usepackage{float} 
\usepackage{geometry}

\begin{document}

\title{A differential model of N player games concerning ethical dilemmas}
\author{
  Ramkrishna Joshi\\
  \textit{Department of Physics, Indian Institute of Technology Hyderabad, India}\\
  \texttt{ph23mscst11029@iith.ac.in} \\
  \and
  Aniruddha Joshi\\
  \textit{Department of Management, PUMBA, Savitribai Phule Pune University, Pune, India}\\
  \texttt{joshiag@gmail.com}
}

\date{\today}

\maketitle

\begin{abstract}
Ethics play an important role in determining the behavior of an individual under certain circumstances. Ethical or unethical behavior can be treated as a strategy of a player in a pay-off game. In this paper, we present two analytical solutions to studying time evolution of behavior of an individual from ethics perspective. We also present the effect of a third player as a perturbation to a two player game and develop a general approach for a N player game. We demonstrate geometric modeling of behavioral characteristics of individuals as polytopes residing in D dimensional space. We treat three player and two player games using set of differential equations that lead to time evolution of phase trajectories which reveal about the interdependencies and self dependencies of each player. We also demonstrate the effect of strategies of each player on other players in cardinal games. 
\end{abstract}

\section{Introduction}

Decision making is a central aspect of Game theory. Outcome of a game is dependent upon the decision matrix of all the players participating in the game. In broad sense, these decisions can be classified as ethical and unethical. In an organizational setting, external factors affecting each of the decision makers determines the time evolution of the organizational behavior\textsuperscript{\href{https://doi.org/10.1007/BF00382936}{1}}. Simplest of the examples of organization behavior is a two player game in which, players Alice and Bob try to maximize their payoffs by choosing certain preferences and strategies\textsuperscript{\href{https://doi.org/10.1038/s41598-023-28627-8}{2},\href{https://press.princeton.edu/books/paperback/9780691130613/theory-of-games-and-economic-behavior?srsltid=AfmBOorLSgPOlm-qjp-RoMPN5Y7Dy3iEJ2zuYf_tx5f7zK9AhIbK6Tkp}{3}}. Key models of this phenomena include zero-sum games and non-zero-sum games that can be effectively studied using Stackelberg strategy\textsuperscript{\href{https://doi.org/10.1007/BF00935665}{13}}. Nash equilibrium represents a state where no player can improve their outcome by unilaterally changing their strategy\textsuperscript{\href{https://www.cs.upc.edu/~ia/nash51.pdf}{4}}.

Traditional cardinal games deal with mapping each players decision to a set of real numbers which can evaluate the effectiveness of the strategies and payoff for each player\textsuperscript{\href{https://doi.org/10.1023/A:1026476425031}{5}}. For such games, interactions between players can be effectively modeled at discrete times with a set of equations. However, modeling the time evolution of these strategies can be a complex problem. In this paper, we provide a simple analytical and geometrical approach to study the continuous time evolution of ethical or unethical behaviors in two player and three player games. 

The most general state of theory asserts that (un)ethical values are gained over lifetime. We prescribe a model to the gain of (un)ethical values and the time evolution of (un)ethical behavioral marker over a period of 100 years of lifetime. 

\begin{equation}
f(\text{Age}) = \frac{1}{1 + e^{\frac{A_0 - \text{Age}}{(TF) \cdot (CF)}}}
\end{equation}
where \( f(\text{Age}) \) represents the behavioral marker, ranging between 0 and 1, as a function of age. The variable \( \text{Age} \) corresponds to the age in years, plotted along the x-axis. The parameter \( A_0 \) is the midpoint age, which determines the age at which the transition occurs. The constant \( TF \), set to 0.1 in this context, acts as a transition factor. The variable \( CF \) referred to as the circumstantial factor controls the sharpness of the transition, with larger values of \( CF \) resulting in a smoother rise.

\begin{figure}[ht]
    \centering
    \begin{minipage}[t]{0.45\textwidth}
        \centering
        \includegraphics[width=\textwidth]{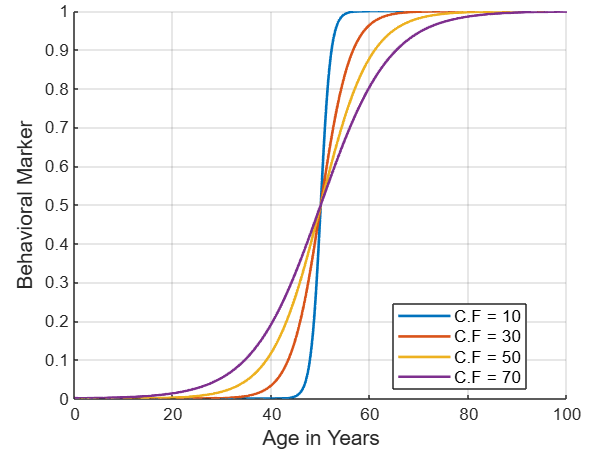} 
        \caption*{(a)}
        \label{fig:ua}
    \end{minipage}
    \hfill
    \begin{minipage}[t]{0.45\textwidth}
        \centering
        \includegraphics[width=\textwidth]{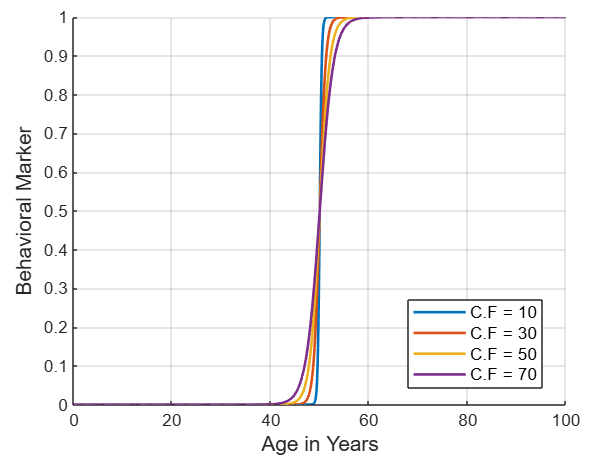} 
        \caption*{(b)}
        \label{fig:eb}
    \end{minipage}
    \caption{The plot represents normalized time evolution of behavioral marker over 100 years with varying circumstantial factors (C.F) for (a)Transition factor=0.1 and (b)Transition factor=0.02, at half age of \( A_0 \)=50 yrs.}
    \label{fig:ua_eb}
\end{figure}

This model is inspired by a realistic idea of the development of (un)ethical values from our childhood. Until a threshold age since birth, an individual acquires no (un)ethical values. This is the part of the graph where behavioral marker is saturated at zero and its neighborhood. Beyond the threshold age, there is gradual development of (un)ethical values and learning that are employed by an individual in daily life. This is the part of the graph where the gradient is non-zero and there is a gradual increase in the behavioral marker. Then another threshold is reached where the behavioral marker saturates near one which we consider to be ceasing period of the (un)ethical development. We assume that this period of the graph represents the complete development of an individual in terms of (un)ethical values and learning gained and that the individual is self sufficient to effectively apply those learning in various scenarios.

Circumstantial factor (CF) is introduced to quantify for all circumstances that influence a person to develop behavioral markers. Higher the value of CF, more gradual is the developmental stage. This can be seen from both the plots. C.F=70 has a greater gradual increase in the behavioral marker compared to other CF values. Transition factor (TF) is introduced to decide the threshold age at which the gradual increase in the behavioral markers begins. As can be readily seen from the plots, higher value of TF accounts for a lesser age at which behavioral markers start to gradually increase.

\subsection{Prisoner's Dilemma (PD)}

The Prisoners’ Dilemma is one of the most well-known concepts in game theory, illustrating the tension between individual rationality and collective benefit. It is a non-zero-sum game that captures the challenges of cooperation and competition, particularly when players act in their own self-interest without communication or trust \textsuperscript{\href{https://doi.org/10.1038/s41598-022-26890-9}{7},\href{http://doi.org/10.1098/rsos.182142}{8},\href{https://doi.org/10.1038/s41598-022-11654-2}{9}}.

Imagine two individuals, Alice and Bob, are arrested for a crime. They are interrogated separately and offered the following options:

\begin{itemize}
\item If Alice and Bob both remain silent (cooperate with each other), they each receive a light sentence of 1 year.
\item If one confesses (defects) while the other remains silent, the defector goes free while the silent partner receives a heavy sentence of 10 years.
\item If both confess (defect), they each receive a moderate sentence of 5 years.
\end{itemize}

Period of sentence is a factor deciding between the cooperation and defecting. The payoff matrix for this scenario is as follows:

\[
\begin{tabular}{|c|c|c|}
\hline
    & \textbf{Bob: Cooperate} & \textbf{Bob: Defect} \\
\hline
\textbf{Alice: Cooperate} & $(1, 1)$ & $(10, 0)$ \\
\hline
\textbf{Alice: Defect} & $(0, 10)$ & $(5, 5)$ \\
\hline
\end{tabular}
\]

In this setup, the lower the number, the better the outcome for the player. The Nash Equilibrium occurs when no player can unilaterally change their strategy to improve their payoff. In the Prisoners’ Dilemma, the Nash Equilibrium is mutual defection (\textquotedblleft Confess\textquotedblright)\textsuperscript{\href{https://press.princeton.edu/books/paperback/9780691130613/theory-of-games-and-economic-behavior?srsltid=AfmBOorLSgPOlm-qjp-RoMPN5Y7Dy3iEJ2zuYf_tx5f7zK9AhIbK6Tkp}{3},\href{https://www.cs.upc.edu/~ia/nash51.pdf}{4},\href{https://doi.org/10.1073/pnas.0308738101}{11}}. Here, neither Alice nor Bob can reduce their sentence by changing their decision, given the other’s strategy. While this is a stable outcome, it is suboptimal because both players would have been better off cooperating.

In this game, the dominant strategy for both players is to defect. Regardless of the other’s choice, defecting minimizes the risk of receiving the worst payoff (10 years in prison). This makes cooperation difficult to sustain, even though mutual cooperation leads to a better collective outcome. In a single-round Prisoners’ Dilemma, defection is the rational choice. However, when the game is repeated over multiple rounds cooperation may emerge as a rational strategy\textsuperscript{\href{https://doi.org/10.1016/0022-2496(78)90030-5}{10}}. 

The Prisoners’ Dilemma highlights the challenges of achieving cooperation in competitive environments. Same can be said about ethical and unethical behavior. In an ethical setting, unethically behaving player can either lead to strong discouragement of unethical behavior or successful manipulation of ethical players into behaving unethically. Similar arguments can be made for unethical setting and ethically behaving player.

\subsection{PD analogue in the theory of ethics}

In this study we develop a classical analogue of the PD scenario characterized by ethical and unethical behavior of players. 

Consider a hypothetical scenario. Two players namely Bob and Alice notice two 100 dollar notes falling loose from the pocket of a well-known businessman in the town. Each one of them picks up one note and have to make one of the following two decisions:

\begin{itemize}

\item Bob/Alice can either keep it to Himself/Herself. He/She will be at the profit of 100 dollars.
\item Bob/Alice can return the note to the businessman and will be awarded 50 dollars in return for their honesty.

\end{itemize}

Given the scenario, one can apply the PD treatment to this problem and generate the payoff matrix as following:

\[
\begin{tabular}{|c|c|c|}
\hline
    & \textbf{Bob: Return} & \textbf{Bob: Keep} \\
\hline
\textbf{Alice: Return} & $(50, 50)$ & $(50, 100)$ \\
\hline
\textbf{Alice: Keep} & $(100, 50)$ & $(100, 100)$ \\
\hline
\end{tabular}
\]
The dominant strategy is of-course to keep the money to themselves for both Bob and Alice as that's the only outcome to maximize profit to each. If now we characterize the behavior of returning the money by Ethical Behavior (EB) and the behavior of keeping the money by Unethical Behavior (UB), then we notice that the profit is maximized by (UB,UB) for Alice and Bob. While the profit is minimized by (EB,EB). In this scenario, one can behave ethically or unethically by choice. However, to maximize the profit, one is better off behaving unethically than ethically.

Consider a slightly modified scenario. The whole businessman setting is a social experiment being performed by a Youtuber. The entire experiment is conducted in exactly same fashion except for one rule.

\textbf{Rule: Whoever returns the money gets to keep the 100 dollar bill and is additionally awarded 50 dollars for their honesty.} 

Thus the new payoff matrix becomes:

\[
\begin{tabular}{|c|c|c|}
\hline
    & \textbf{Bob: Return} & \textbf{Bob: Keep} \\
\hline
\textbf{Alice: Return} & $(150, 150)$ & $(150, 100)$ \\
\hline
\textbf{Alice: Keep} & $(100, 150)$ & $(100, 100)$ \\
\hline
\end{tabular}
\]

With this new payoff matrix, (EB,EB) is the new way to maximize profit for both. We refer to this scenario as Forced Ethical Behavior (FEB) since due to the additional payoff, each player is forced to behave ethically to maximize the profit. We claim that though FEB scenario is realistic, players cannot be forced to behave ethically without the true prior information of the social experiment and hence this scenario can be used as a mesh to separate truly ethical individuals from unethical individuals. Truly ethical individuals are defined to be the individuals who without the prior knowledge on the experiment will choose to behave ethically and return the money. In this context we also define Conditionally ethical individuals. Conditionally ethical individuals are the ones who would choose to behave ethically while being informed of the true prior information of the experiment. These individuals, without any prior information will choose to keep the money to maximize their profit. And the last category of individuals are truly unethical individuals. These individuals will choose to keep the money even after gaining true information of the experiment. 

Thus apart from truly ethical individuals, no other category of individuals will behave ethically unless an extra incentive is applicable for ethical behavior.

A much simpler matrix can be prepared to evaluate the situation. Given normalizable ethical and unethical behavior, we describe two parameters namely; $0 \leq E \leq 1$ and  $0 \leq UE \leq 1$ where the former represents the ethical behavioral parameter and the latter represents the unethical behavioral parameter. Both these parameters are related to each other as; 

\begin{equation}
E+UE = 1
\end{equation}

Hence in a two-player game, the matrix is given by;

\[
\begin{matrix}
    & E & \text{UE} \\
    \text{Alice} & ( \phi_{11}, & \phi_{12} ) \\ \\
    \text{Bob} & ( \phi_{21}, & \phi_{22}) \\ \\
\end{matrix}
\]

where, $0 \leq \phi_{11},\phi_{12},\phi_{21},\phi_{22} \leq 1$. The relations between the matrix elements are;

\begin{equation}
\phi_{12} = 1-\phi_{11},\
\phi_{22} = 1-\phi_{21}
\end{equation}

Any behavioral state of Alice or Bob can be obtained by a convex mixture of the extrema states. The above set forms a convex set for Alice and Bob with state vectors E and UE. To find the eigenvalues of matrix, we solve the characteristic equation:
\[
\det(\Phi - \lambda I) = 0,
\]
where \( I \) is the \( 2 \times 2 \) identity matrix, and \( \lambda \) is the eigenvalue. Substituting \( \Phi - \lambda I \):
\[
\Phi - \lambda I = 
\begin{pmatrix}
\phi_{11} - \lambda & \phi_{12} \\
\phi_{21} & \phi_{22} - \lambda
\end{pmatrix}.
\]

The determinant is:
\[
\det(\Phi - \lambda I) = (\phi_{11} - \lambda)(\phi_{22} - \lambda) - \phi_{12} \phi_{21}.
\]

On performing algebra and substituting, it can be shown that;
\[
\det(\Phi - \lambda I) = \lambda^2 - (\phi_{11} + \phi_{21})\lambda + \phi_{11} - \phi_{21}.
\]

Solutions to quadratic equation are;
\[
\lambda = \frac{\phi_{11} + \phi_{21} \pm \sqrt{(\phi_{11} + \phi_{21})^2 - 4(\phi_{11} - \phi_{21})}}{2}.
\]

For each eigenvalue \( \lambda \), the eigenvector \( \mathbf{v} \) satisfies:
\[
(\Phi - \lambda I)\mathbf{v} = 0.
\]

and hence can be easily calculated to be;
\[
\mathbf{v}_1 = \begin{pmatrix}
-\frac{1 - \phi_{11}}{\phi_{11} - \lambda_1} \\
1
\end{pmatrix}
\]
\[
\mathbf{v}_2 = \begin{pmatrix}
-\frac{1 - \phi_{11}}{\phi_{11} - \lambda_2} \\
1
\end{pmatrix}
\]

Respectively for; 

\begin{equation}
\lambda_{1,2} = \frac{\phi_{11} + \phi_{21} \pm \sqrt{(\phi_{11} + \phi_{21})^2 - 4(\phi_{11} - \phi_{21})}}{2}.
\end{equation}

Consider a special case where Alice is a true ethical player and Bob is a true unethical player. The behavioral matrix in such a scenarios is identity matrix (I) since $\phi_{11}=\phi_{22}=1$. We assume a generic time evolution of the identity matrix under influence of external parameters for both Alice and Bob.

\begin{figure}[H]
    \centering
    \includegraphics[width=0.5\linewidth]{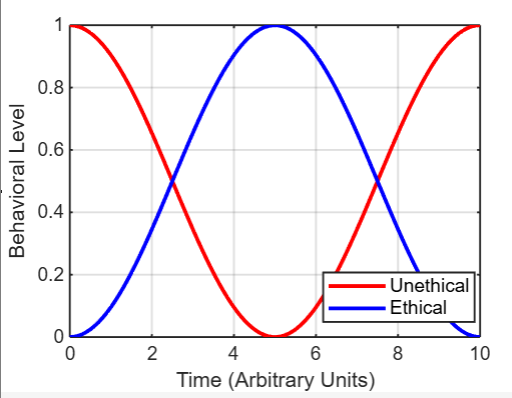}
    \caption{Time evolution of behavioral parameters of Alice and Bob in the simple $\phi$ matrix scheme.}
    \label{fig:enter-label}
\end{figure}

The nature of the evolution is one unique solution to the constrained equations. A continuous and differentiable solution simplifies the analysis. An abrupt change in the behavioral marker can be challenging to explain in terms of the time evolution matrix.However, one is free to choose any required prior on the distribution. Different priors give different posteriors.

One is ultimately interested in understanding behavior of a third party/organization (C) under the influence of the behavior of Alice and Bob. This problem involves understanding 8 different probabilities, namely; $P(E_C|E_A,E_B),P(E_C|E_A,U_B),P(E_C|U_A,E_B),P(E_C|U_A,U_B)$ and similar four combinations for $U_C$ which include $P(U_C|E_A,E_B),P(U_C|E_A,U_B),P(U_C|U_A,E_B),P(U_C|U_A,U_B)$. For simplicity, by assuming the independence of behavioral outcomes of Alice and Bob, the boolean truth table of this event includes 16 probability vectors. In organizational setting, $E_C$ and $U_C$ can be considered to be first order perturbation of a third player in a two player game and can have wide ranging effects on dynamics of the game. It is also beneficial to define the magnitude of the effect of this perturbation on the game.

\begin{center}
\begin{tabular}{|c|c|c|c||c|c|c|c|}
\hline
$U_A$ & $E_A$ & $U_B$ & $E_B$ & $U_A \cap E_A$ & $U_A \cap U_B$ & $U_A \cap E_B$ & $E_A \cap E_B$ \\
\hline
0 & 0 & 0 & 0 & 0 & 0 & 0 & 0 \\
0 & 0 & 0 & 1 & 0 & 0 & 0 & 0 \\
0 & 0 & 1 & 0 & 0 & 0 & 0 & 0 \\
0 & 0 & 1 & 1 & 0 & 0 & 0 & 0 \\
0 & 1 & 0 & 0 & 0 & 0 & 0 & 0 \\
0 & 1 & 0 & 1 & 0 & 0 & 0 & 1 \\
0 & 1 & 1 & 0 & 0 & 0 & 0 & 0 \\
0 & 1 & 1 & 1 & 0 & 0 & 0 & 1 \\
1 & 0 & 0 & 0 & 0 & 0 & 0 & 0 \\
1 & 0 & 0 & 1 & 0 & 0 & 1 & 0 \\
1 & 0 & 1 & 0 & 0 & 1 & 0 & 0 \\
1 & 0 & 1 & 1 & 0 & 1 & 1 & 0 \\
1 & 1 & 0 & 0 & 1 & 0 & 0 & 0 \\
1 & 1 & 0 & 1 & 1 & 0 & 1 & 1 \\
1 & 1 & 1 & 0 & 1 & 1 & 0 & 0 \\
1 & 1 & 1 & 1 & 1 & 1 & 1 & 1 \\
\hline
\end{tabular}
\end{center}

Considering that $U_B+E_B=1$ and $U_A+U_B=1$, we restrict the truth table to only four entries by considering extremal events. This is since if Bob is behaving ethically with 1 probability then he must behave unethically with 0 probability and vice versa. Same applies for Alice.

\begin{center}
\begin{tabular}{|c|c|c|c||c|c|c|c|}
\hline
$U_A$ & $E_A$ & $U_B$ & $E_B$ & $U_A \cap E_A$ & $U_A \cap U_B$ & $U_A \cap E_B$ & $E_A \cap E_B$ \\
\hline
0 & 1 & 0 & 1 & 0 & 0 & 0 & 1 \\
0 & 1 & 1 & 0 & 0 & 0 & 0 & 0 \\
1 & 0 & 0 & 1 & 0 & 0 & 1 & 0 \\
1 & 0 & 1 & 0 & 0 & 1 & 0 & 0 \\
\hline
\end{tabular}
\end{center}

To further simplify the analysis, we only consider the subset of events where we consider only ethical markers and ignore unethical marker. This is justified since we are only looking at extremal events.

\begin{center}
\begin{tabular}{|c|c||c|}
\hline
$E_A$ & $E_B$ & $P(E_C \mid E_A, E_B)$ \\
\hline
0 & 0 & 0 \\
0 & 1 & 0 \\
1 & 0 & 0 \\
1 & 1 & 1 \\
\hline
\end{tabular}
\end{center}

This suggests that given any one of the players is behaving unethically, the organization is bound to behave unethically. This can be geometrically understood by constructing a 3 dimensional polytope with these four vertices. 

A polytope in \( \mathbb{R}^n \) is a convex, compact set that can be defined as the intersection of a finite number of half-spaces. More formally, a polytope is the convex hull of a finite set of points, or equivalently, the solution set to a finite system of linear inequalities. Let \( P \subset \mathbb{R}^n \) be a polytope, then there exist vectors \( \mathbf{a}_1, \mathbf{a}_2, \dots, \mathbf{a}_m \in \mathbb{R}^n \) and scalars \( b_1, b_2, \dots, b_m \in \mathbb{R} \) such that:

\[
P = \left\{ \mathbf{x} \in \mathbb{R}^n \mid \mathbf{a}_i^T \mathbf{x} \leq b_i \text{ for all } i = 1, 2, \dots, m \right\}.
\]

Here, the vectors \( \mathbf{a}_i \) define the normal directions to the hyperplanes that form the boundary of the polytope, and the inequalities \( \mathbf{a}_i^T \mathbf{x} \leq b_i \) describe the half-spaces that intersect to form the polytope. 

Polytopes are characterized by their vertices (the extreme points), edges, and faces. A polytope is called a convex polytope if for any two points within the polytope, the line segment joining them also lies entirely within the polytope. For this setup, we generate the polytopes using the following set of inequalities;

\begin{equation}
0 \leq E_A \leq 1
\end{equation}

\begin{equation}
0 \leq E_B \leq 1
\end{equation}

\begin{equation}
0 \leq P(E_C | E_A, E_B) \leq 1
\end{equation}

\begin{equation}
P(E_C | E_A, E_B) \geq E_A
\end{equation}

\begin{equation}
P(E_C | E_A, E_B) \geq E_B
\end{equation}

\begin{equation}
P(E_C | E_A, E_B) \leq E_A + E_B
\end{equation}

We refer to the generated polytope here as convex ethical polytope since it is a result of the ethical perturbation with initial state of both players (Alice and Bob) being ethical. In a similar fashion, one can also define an convex unethical polytope which is a result of unethical initial state of both players. Two more convex partially ethical polytopes can be generated where the initial state is characterized by one player being ethical and the other being unethical.

\begin{figure}[H]
    \centering
    \includegraphics[width=0.5\linewidth]{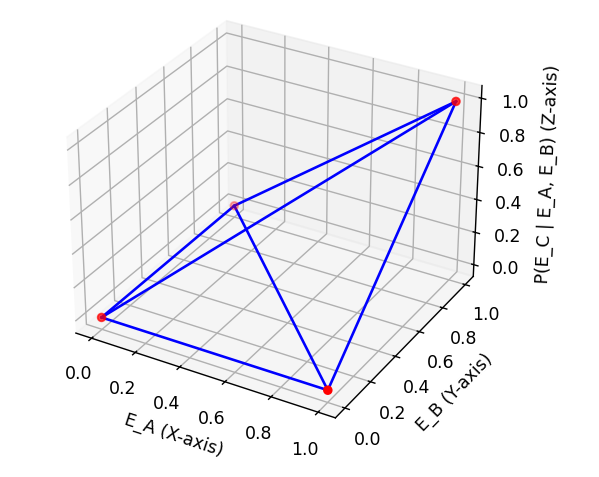}
    \caption{An ethical polytope obtained with four vertices representing four independent extremal events for ethical behavior of Alice, Bob and the third party C with AND condition.}
    \label{fig:enter-label}
\end{figure}

Similar polytopes can be generated for the remaining seven probability vectors. Every possible permutations and combinations of ethical behaviors of Alice, Bob and party C are included in the interior of the above polytope. Any event inside the polytope can be expressed as a convex combination of the basis vectors i.e x,y and z coordinates of the polytope. Similar to the intersection of events (AND gate), union of events can also be generated (OR gate).

\begin{center}
\begin{tabular}{|c|c|c|c||c|c|c|c|}
\hline
$U_A$ & $E_A$ & $U_B$ & $E_B$ & $U_A \cup E_A$ & $U_A \cup U_B$ & $U_A \cup E_B$ & $E_A \cup E_B$ \\
\hline
0 & 1 & 0 & 1 & 1 & 0 & 1 & 1 \\
0 & 1 & 1 & 0 & 1 & 1 & 0 & 1 \\
1 & 0 & 1 & 0 & 1 & 1 & 1 & 0 \\
1 & 0 & 1 & 0 & 1 & 1 & 1 & 0 \\
\hline
\end{tabular}
\end{center}

From this table, another polytope can be generated with the following vertices;

\begin{center}
\begin{tabular}{|c|c||c|}
\hline
$E_A$ & $E_B$ & $P(E_C | E_A, E_B)$ \\
\hline
0 & 0 & 0 \\
0 & 1 & 1 \\
1 & 0 & 1 \\
1 & 1 & 1 \\
\hline
\end{tabular}
\end{center}

\begin{figure}[H]
    \centering
    \includegraphics[width=0.5\linewidth]{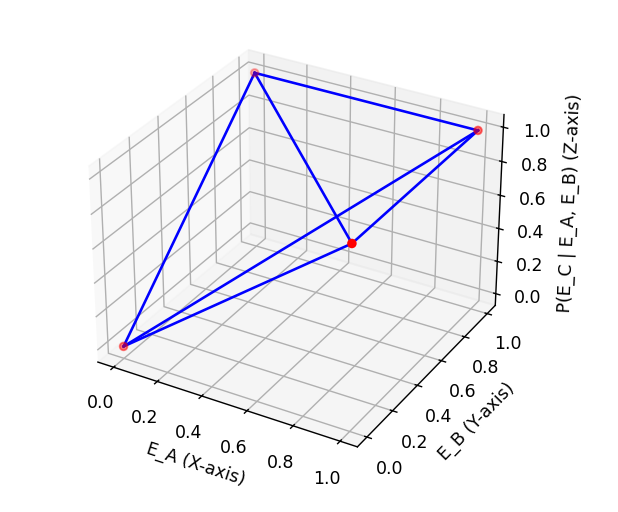}
    \caption{An ethical polytope obtained with four vertices representing four independent extremal events for ethical behavior of Alice, Bob and the third party C with OR condition.}
    \label{fig:enter-label}
\end{figure}

Combining the above two polytopes gives us a complete picture of organizational behavior under influence of the behavior of Alice and Bob.

\begin{figure}[H]
    \centering
    \includegraphics[width=0.5\linewidth]{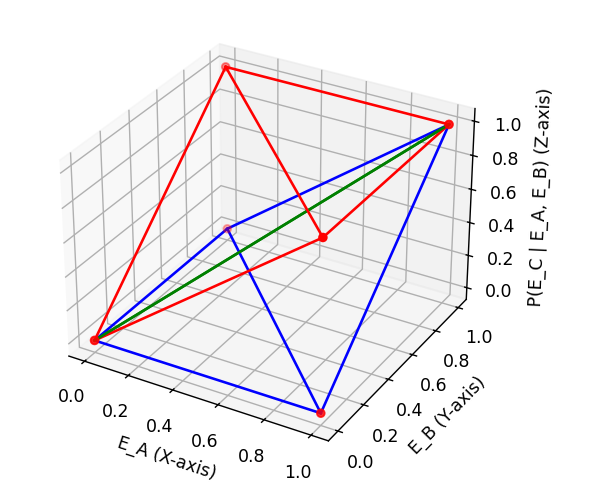}
    \caption{Combined ethical polytope of AND and OR conditions. AND condition polytope is represented with blue edges, OR condition polytope is represented with red edges and common edges are represented with green.}
    \label{fig:enter-label}
\end{figure}

The common solutions for the AND and OR conditions include (0,0,0) and (1,1,1) i.e when all three Alice, Bob and C behave either unethically or ethically. These solutions are in good agreement with our experience that if two parties behave (un)ethically then the third party C,is also almost always likely to behave (un)ethically. "Almost" is used here to indicate of any perturbations that may affect the behavior of C party otherwise. In a similar way, polytopes can be used as a geometric tool to generate and understand all other scenarios.

\section{Differential approach to the theory of ethics: Two player games}

\subsection{Love dynamics: A mathematical mode}

Steven Strogatz has presented an extensive work on mathematical modeling of love affair between two individuals Romeo and Juliet \textsuperscript{\href{https://www.biodyn.ro/course/literatura/Nonlinear_Dynamics_and_Chaos_2018_Steven_H._Strogatz.pdf}{6}}. Their love and hatred towards each other can be effectively modeled as a function of time by using a system of coupled differential equations. This system of equations models the interaction of Romeo and Juliet through their individual as well as mutual feelings for each other.

Let \( R(t) \) be the emotional state of Romeo at time \( t \), and \( J(t) \) be the emotional state of Juliet at the same time. The interaction between Romeo and Juliet's emotions can be described by a system of coupled differential equations.

\[
\frac{dR}{dt} = \alpha_1 J + \beta_1 R
\]
\[
\frac{dJ}{dt} = \alpha_2 R + \beta_2 J
\]

where \( \alpha_1 \)/\( \beta_1 \) are constants that represent the rate at which Juliet's/Romeo's feelings influence Romeo. While \( \alpha_2 \)/\( \beta_2 \) are constants that represent the rate at which Romeo's/Juliet's feelings influence Juliet.
The values of the parameters \( \alpha_1, \alpha_2, \beta_1, \beta_2 \) determine the nature of the relationship between Romeo and Juliet.
The parameters \( \alpha_1 \) and \( \alpha_2 \) represent dynamics of interpersonal relations while parameters \( \beta_1 \) and \( \beta_2 \) represent self feelings. In general, positive values of the parameters represent a attractive influence of feelings and negative values represent repulsive influence of feelings.\( \alpha_1 > 0 \) means that Juliet’s love has a positive influence on Romeo, i.e., the more Juliet loves Romeo, the more Romeo's love for her increases while \( \alpha_1 < 0 \) means that Juliet’s love has a negative influence on Romeo, i.e., Romeo’s love decreases as Juliet's love for him increases. This could occur in a situation where Romeo feels overwhelmed or repelled by Juliet's affection. Similarly, \( \alpha_2 > 0 \) means that Romeo's love has a positive influence on Juliet, i.e., the more Romeo loves Juliet, the more Juliet's love for him increases while \( \alpha_2 < 0 \) means that Romeo's love has a negative influence on Juliet, i.e., Juliet’s love decreases as Romeo's love for her increases. This could happen in a relationship where Juliet feels stifled by Romeo’s affection. The parameters \( \beta_1 \) and \( \beta_2 \) represent the natural decay or fading of love when there is no reciprocal affection. \( \beta_1 < 0 \) means that Romeo’s love decays over time if Juliet does not reciprocate, i.e., without Juliet’s affection, Romeo’s emotional state will gradually diminish. \( \beta_1 > 0 \) means that Romeo’s love will grow over time even in the absence of Juliet’s affection, suggesting an obsessive or one-sided attachment.\( \beta_2 < 0 \) means that Juliet’s love decays over time if Romeo does not reciprocate, i.e., without Romeo’s affection, Juliet’s emotional state will gradually diminish and \( \beta_2 > 0 \) means that Juliet’s love will grow over time even in the absence of Romeo’s affection, indicating a potentially obsessive or unrequited affection.

Several special cases highlighted by Strogatz in [12] include:

\begin{itemize}

    \item Eager Beaver: When Romeo is excited by his own feeling towards Juliet as well as by her feelings towards him this case is specified by \( \beta_1 > 0 \) and \( \alpha_1 > 0 \).
    \item Narcissistic Nerd: When Romeo is excited by his own feelings towards Juliet but is put off by her feelings towards him, his situation is characterized by \( \beta_1 < 0 \) and \( \alpha_1 > 0 \).
    \item Cautious lover: When Romeo is excited about the feelings of Juliet towards him but holds himself back because of his own feelings, he is characterized by \( \beta_1 > 0 \) and \( \alpha_1 < 0 \).
    \item Hermite: When Romeo is not excited by feelings of Juliet nor by his own feelings, both \( \beta_1 < 0 \) and \( \alpha_1 < 0 \).
    
\end{itemize}

Based upon above prescribed scenarios,the system of equations can exhibit several types of behavior depending on the values of the parameters. In conclusion,the dynamics of love between Romeo and Juliet can be effectively modeled using differential equations that capture the interaction between their emotional states. The parameters \( \alpha_1, \alpha_2, \beta_1, \beta_2 \) provide insight into how each person's emotions are influenced by the other and how they evolve over time. By adjusting these parameters, we can model various types of relationships, from stable love to oscillations and even chaos.

\subsection{Behavioral dynamics: Ethical and Unethical behavior}

The dynamics of ethical and unethical behavior can also be effectively modeled using a system of differential equations. In this model, we define the ethical behavior of an individual as \( E(t) \) and their unethical behavior as \( U(t) \), where \( t \) represents time. The evolution of these behaviors is influenced not only by their own inherent properties but also by the interaction between the two. This interaction is described by the following system of coupled differential equations:

\[
\frac{dE}{dt} = \alpha_1 E + \beta_1 U
\]
\[
\frac{dU}{dt} = \alpha_2 E + \beta_2 U
\]

Here, \( \alpha_1 \) represents the rate at which ethical behavior influences ethical behavior, \( \alpha_2 \) denotes the influence of ethical behavior on unethical behavior, \( \beta_1 \) reflects the influence of unethical behavior on ethical behavior, and \( \beta_2 \) represents the influence of unethical behavior on unethical behavior.

The parameters \( \beta_1 \) and \( \alpha_2 \) describe the reciprocal relationship between ethical and unethical behaviors. A positive value of \( \beta_1 \) implies that unethical behavior has a positive influence on ethical behavior. This might occur, for example, when witnessing or confronting unethical actions leads a person to reaffirm or adopt more ethical standards. Conversely, a negative value of \( \beta_1 \) suggests that unethical behavior detracts from ethical behavior, which might happen in environments where unethical actions weaken or undermine the ethical standards of individuals. Similarly, a positive value of \( \alpha_2 \) means that ethical behavior encourages unethical behavior, perhaps in cases where ethical decisions lead to unintended consequences, such as enabling or inadvertently promoting unethical actions. A negative value of \( \alpha_2 \) indicates that ethical behavior works to suppress unethical behavior, implying that the presence of ethical actions in a person or community reduces the occurrence of unethical behavior.

Parameters \( \alpha_1 \) and \( \beta_2 \) capture the self coupling of ethical and unethical behaviors over time when they are not reinforced. A negative value of \( \alpha_1 \) suggests that ethical behavior decreases over time without reinforcement, highlighting that ethical actions need continuous support to be sustained. If \( \alpha_1 \) is positive, it implies that ethical behavior strengthens or becomes more ingrained over time, even in the absence of external encouragement. Similarly, a negative value of \( \beta_2 \) indicates that unethical behavior disappears without reinforcement, while a positive value of \( \beta_2 \) suggests that unethical behavior grows or becomes more pronounced over time if not checked.

The interaction between ethical and unethical behaviors in this system can give rise to different types of dynamic behavior, depending on the values of the parameters. In scenarios where \( \beta_1, \alpha_2 < 0 \) and \( \alpha_1, \beta_2 > 0 \), the system may settle into a stable equilibrium where both ethical and unethical behaviors stabilize at constant values. This represents a situation where the reinforcement forces between the two behaviors lead to a balance.

However, if the parameters \( \beta_1 \) and \( \alpha_2 \) have opposite signs, the system may exhibit oscillatory dynamics. For example, if \( \beta_1 > 0 \) and \( \alpha_2 < 0 \), unethical behavior could initially increase ethical behavior, but over time, ethical behavior would counteract unethical behavior, leading to periodic cycles of increase and decrease in both behaviors. This oscillatory nature reflects the dynamic tension between ethical and unethical tendencies in a person or group.

In more complex cases, where the parameters \( \alpha_1, \alpha_2, \beta_1, \beta_2 \) are set such that the system exhibits sensitivity to initial conditions, chaotic behavior may emerge. This occurs when small changes in initial ethical or unethical behaviors lead to large, unpredictable shifts in the system's trajectory. Such chaotic behavior suggests that the dynamics of ethical and unethical tendencies are highly sensitive to both external and internal factors, and the resulting patterns of behavior are difficult to predict.

In conclusion, this model provides a mathematical framework for understanding the dynamic relationship between ethical and unethical behavior. The interactions between these behaviors, governed by the parameters \( \alpha_1, \alpha_2, \beta_1, \beta_2 \), reveal that ethical and unethical tendencies can both influence each other in complex ways, with the possibility of stable equilibria, oscillations, or chaotic dynamics. The system highlights the importance of reinforcement and decay in shaping moral behavior, providing valuable insights into the ways in which ethical and unethical actions evolve and interact over time.

Above scenario can be further bifurcated from a viewpoint of a two-player game. Once again, consider that Bob and Alice are two players participating in the game. This game can have four possible initial conditions, namely;

\begin{itemize}
    \item Alice is an unethical player and Bob is an ethical player at the start of the game.
    \item Bob is an unethical player and Alice is an ethical player at the start of the game.
    \item Both Alice and Bob are ethical players at the start of the game.
    \item Both Alice and Bob are unethical players at the start of the game.
\end{itemize}

Let the ethical markers for Bob and Alice be $E_B$ and $E_A$ respectively and unethical markers be $U_A$ and $U_B$. Thus, four initial conditioned games can be mathematically modeled as;

\begin{itemize}
\item 1. Ethical Bob and Crook Alice
\[
\frac{dU_A}{dt} = \alpha_1 U_A + \beta_1 E_B
\]
\[
\frac{dE_B}{dt} = \alpha_2 U_A + \beta_2 E_B
\]

\item 2. Ethical Alice and Crook Bob
\[
\frac{dU_B}{dt} = \alpha_1 U_B + \beta_1 E_A
\]
\[
\frac{dE_A}{dt} = \alpha_2 U_B + \beta_2 E_A
\]
\end{itemize}
Above two situations represent the interaction of an ethical and an unethical person. Since case 1 and case 2 are exactly identical to each other, we only analyze case 1 and allow the reader to analyze case 2 in the same fashion. A simplified scenario can be readily derived from the above set of equation by letting $\beta_1$=$\alpha_2$=0 and $\beta_1,\beta_2 < 0$. This simplification reduces differential equations to the following form:
\[
\frac{dU_A}{dt} = |\alpha_1| U_A
\]
\[
\frac{dE_B}{dt} = |\beta_2| E_B
\]

Solutions to this set of differential equations is given by, $U_A=e^{|\alpha_1|t}$ and $E_B=e^{|\beta_2|t}$. The (un)ethical behavior of a player will exponentially increase/decrease over time without the influence of the counterpart. Positive $\beta$ values indicate an exponential increase and negative values indicate an exponential decay. Thus an (un)ethical player who is attracted by his/her behavior will continue to behave (un)ethically forever and (un)ethical player who is repelled by his/her behavior will eventually cease to behave (un)ethically.

\begin{figure}[ht]
    \centering
    \begin{minipage}[t]{0.45\textwidth}
        \centering
        \includegraphics[width=\textwidth]{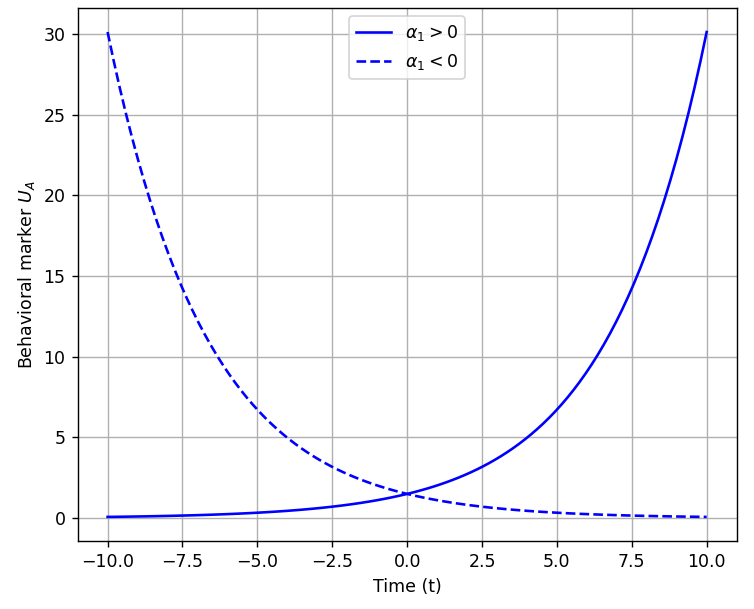} 
        \caption*{(a)}
        \label{fig:ua}
    \end{minipage}
    \hfill
    \begin{minipage}[t]{0.45\textwidth}
        \centering
        \includegraphics[width=\textwidth]{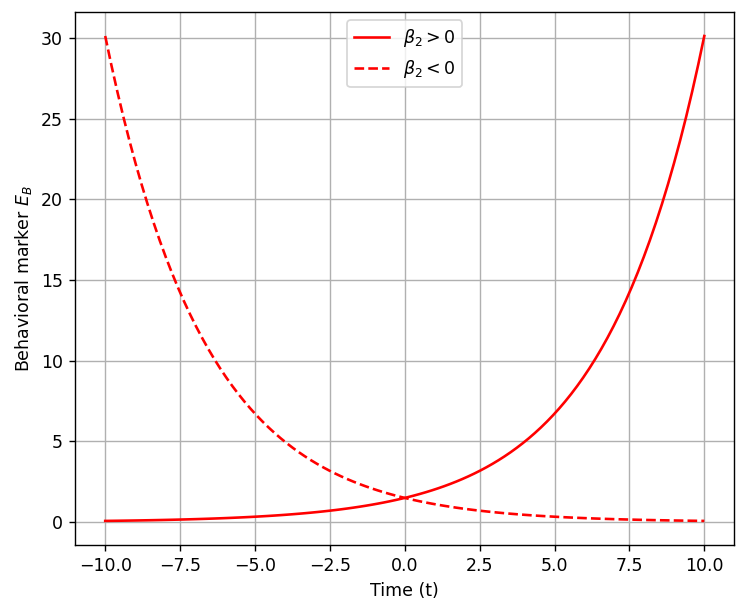} 
        \caption*{(b)}
        \label{fig:eb}
    \end{minipage}
    \caption{Non-normalized behavioral markers: (a) $U_A$ and (b) $E_B$. Solid lines corresponding to $\alpha_{1}>0,\beta_{1}>0$ represent exponential increase and dotted lines corresponding to $\alpha_{1}<0,\beta_{1}<0$ represent the exponential decrease in the ethical behavior of Bob and unethical behavior of Alice . For positive parameters, initial behavioral level is kept at $U_A(t=0), E_B(t=0) = 0.01$. For negative parameters, initial behavioral level is kept at $U_A(t=0), E_B(t=0) = 1.5$.}
    \label{fig:ua_eb}
\end{figure}

We explore a special well known case from analysis by Strogatz, where parametrization is given by $\beta_1=-\alpha_2$ and $\beta_2=-\alpha_1$. We slightly modify the parametrization which modifies the differential equations to,

\[
\frac{dU_A}{dt} = \alpha_1 U_A + \beta_1 E_B
\]
\[
\frac{dE_B}{dt} = \alpha_2 U_A + \beta_2 E_B
\]

This problem can be simply solved by considering a matrix $\psi$ defined as;

\[
\psi = 
\begin{bmatrix}
\alpha_1 & \beta_1 \\
-\beta_1 & -\alpha_1
\end{bmatrix}
\]

Eigenvalues of this matrix can be readily found by using characteristic equation.

\[
\det(\psi - \lambda I) = 0,
\]
where $I$ is the identity matrix, and $\lambda$ is the eigenvalue. Substituting $\psi$, we have:
\[
\psi - \lambda I = 
\begin{bmatrix}
\alpha_1 - \lambda & \beta_1 \\
-\beta_1 & -\alpha_1 - \lambda
\end{bmatrix}.
\]

The determinant is:
\[
\det(\psi - \lambda I) = 
\begin{vmatrix}
\alpha_1 - \lambda & \beta_1 \\
-\beta_1 & -\alpha_1 - \lambda
\end{vmatrix}.
\]

On expanding the determinant and simplifying;
\[
\det(\psi - \lambda I) = (\alpha_1 - \lambda)(-\alpha_1 - \lambda) - (-\beta_1)(\beta_1).
\]

\[
\det(\psi - \lambda I) = -\alpha_1^2 + \lambda^2 + \beta_1^2.
\]

Substitute $\beta_1 = -\alpha_2$, so $\beta_1^2 = \alpha_2^2$:
\[
\det(\psi - \lambda I) = \lambda^2 - (\alpha_1^2 + \alpha_2^2) = 0.
\]

\[
\therefore
\lambda = \pm \sqrt{\alpha_1^2 - \alpha_2^2}.
\]

Thus, the eigenvectors are:
\[
\mathbf{v}_+ = \begin{bmatrix}
1 \\
-\frac{\alpha_1 - \sqrt{\alpha_1^2 - \alpha_2^2}}{\beta_1}
\end{bmatrix},
\quad
\mathbf{v}_- = \begin{bmatrix}
1 \\
-\frac{\alpha_1 + \sqrt{\alpha_1^2 - \alpha_2^2}}{\beta_1}
\end{bmatrix}.
\]

In this section of the paper, we study above set of solutions in a case wise basis.
\begin{itemize}

\item Case 1: Consider a case where $|\alpha_{1}|>|\beta_{1}|$ which implies that $-\sqrt{\alpha_1^2 - \alpha_2^2}<0<\sqrt{\alpha_1^2 - \alpha_2^2}$. Time evolution for this condition is given by;

\[
\begin{bmatrix}
A(t) \\
B(t)
\end{bmatrix}
= C_1 v_{+} e^{\sqrt{\alpha_1^2 - \alpha_2^2}\, t} + C_2 v_{-} e^{-\sqrt{\alpha_1^2 - \alpha_2^2}\, t}
\]

Where A and B are respective behavioral markers of Alice and Bob such that $A \in (U_{A},E_{A})$ and $B \in (U_{B},E_{B})$ and $C_1,C_2$ are normalization constants. 

\item Case 1.1 : Let the first subcase be that $\alpha_{1}>0,\beta_{1}>0$. This means that Alice is a Eager Beaver and Bob is a Hermite. Bob reacts negatively to his own as well as Alice's behavior and Alice reacts positively to hers and Bob's behavior. Given that $|\alpha_{1}|>|\beta_{1}|$, the relative difference between the two changes phase portraits to a great extent. If there is relatively larger separation between the two parameters, then we notice that the the  phase trajectories spread out about the unit slope line cutting the second and the fourth quadrants. On the contrary, if the relative separation between them is smaller, the trajectories squeeze about the same line. This is evident in Figure 7.

\begin{figure}[ht]
    \centering
    \begin{minipage}[t]{0.45\textwidth}
        \centering
        \includegraphics[width=\textwidth]{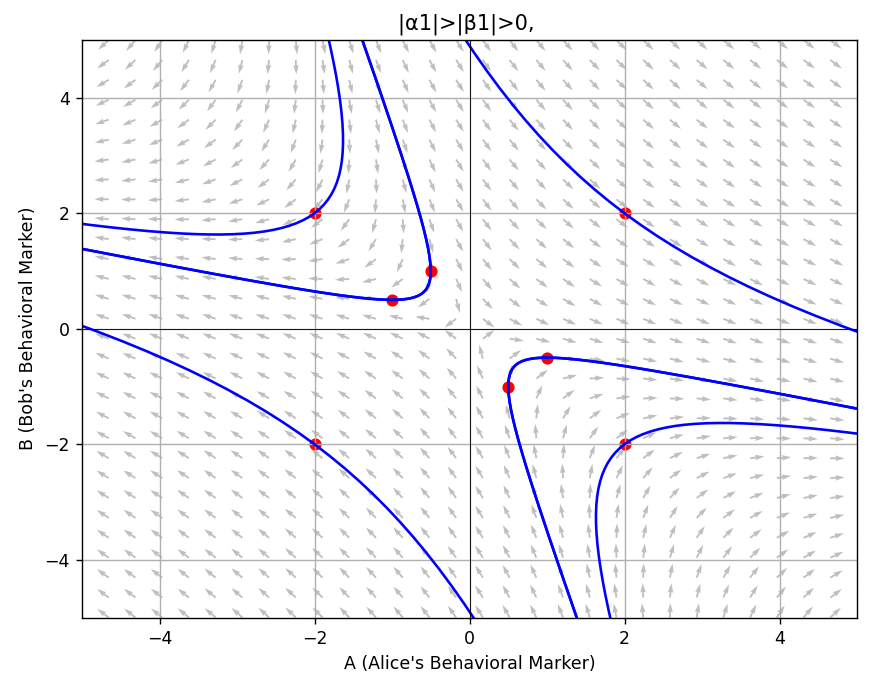} 
        \caption*{(a)}
        \label{fig:ua}
    \end{minipage}
    \hfill
    \begin{minipage}[t]{0.45\textwidth}
        \centering
        \includegraphics[width=\textwidth]{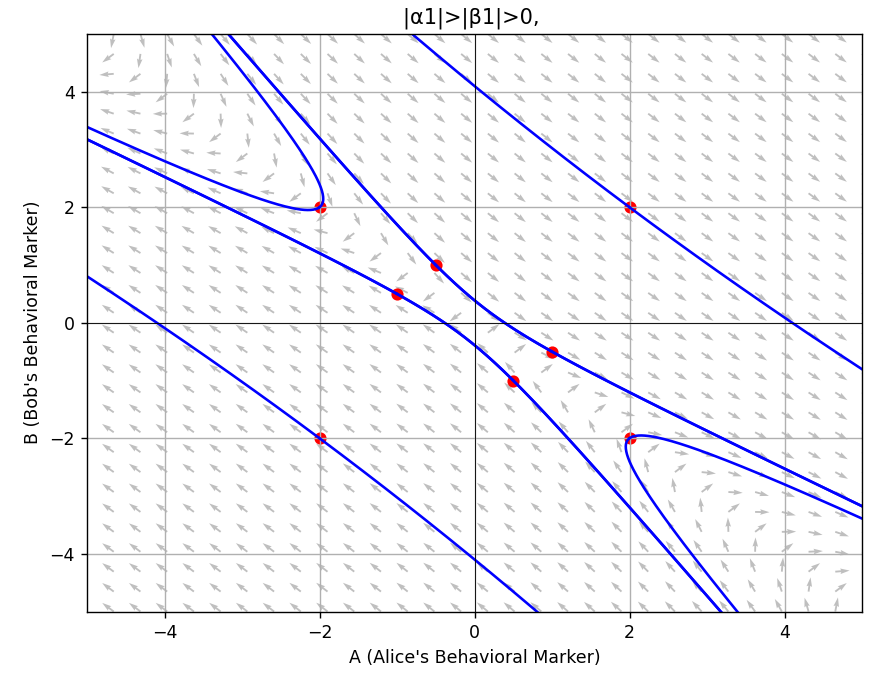} 
        \caption*{(b)}
        \label{fig:eb}
    \end{minipage}
    \caption{Phase portrait of time evolution of behavioral markers for Alice and Bob when (a)$\alpha_{1}=2,\beta_{1}=1,\alpha_{2}=-1,\beta_{2}=-2$ and (b)$\alpha_{1}=1.1,\beta_{1}=1,\alpha_{2}=-1,\beta_{2}=-1.1$ for eight initial conditions of (-2, -2), (2, 2), (-2, 2), (2, -2), (0.5, -1), (-1, 0.5), (1, -0.5), (-0.5, 1) }
    \label{fig:ua_eb}
\end{figure}

Since Alice is an unethical player and Bob is an ethical player at the start of the game, Alice reacts positively to her own unethical behavior as well as the ethical behavior of Bob. Bob however, reacts negatively to his own ethical behavior as well as Alice's unethical behavior depending upon the sign of parameters. The opposite scenario can also be generated where Alice reacts negatively globally and Bob reacts positively globally. This can be seen from the phase portrait. In the second quadrant, as the Alice reacts more and more positively to the ethical behavior of Bob, Bob reacts more and more negatively to Alice's unethical behavior. Similarly, in the fourth quadrant, as Bob reacts more and more positively to the Alice's unethical behavior, Alice reacts more negatively to Bob's ethical behavior. Response of behaviors of Alice and Bob is exactly contrary to each other. 

\item Case 1.2 : Let the second subcase be that $\alpha_{1}>0,\beta_{1}<0$. This means that Alice is a Narcissistic nerd and Bob is a cautious lover. Bob reacts negatively to his own behavior and positively to Alice's behavior. Alice reacts positively to her own behavior but negatively to Bob's behavior.Two dynamic scenarios can arise in this context. An ethical player is highly susceptible to being influenced by the unethical atmosphere around him/her. Once the unethical influence grows over the player, he/she is repelled by his/her own ethical tendencies. In this scenario, where the unethical player dominates, he/she is highly repelled by the ethical tendencies of the ethical players.Alice is a narcissistic nerd in this scenario. Hence she reacts positively to her own unethical behavior and negatively to the ethical behavior of Bob. Bob reacts negatively to his own ethical behavior and positively to the unethical behavior of Alice. Hence, this scenario represents the unethical influence of Alice on Bob.

\begin{figure}[H]
    \centering
    \begin{minipage}[t]{0.45\textwidth}
        \centering
        \includegraphics[width=\textwidth]{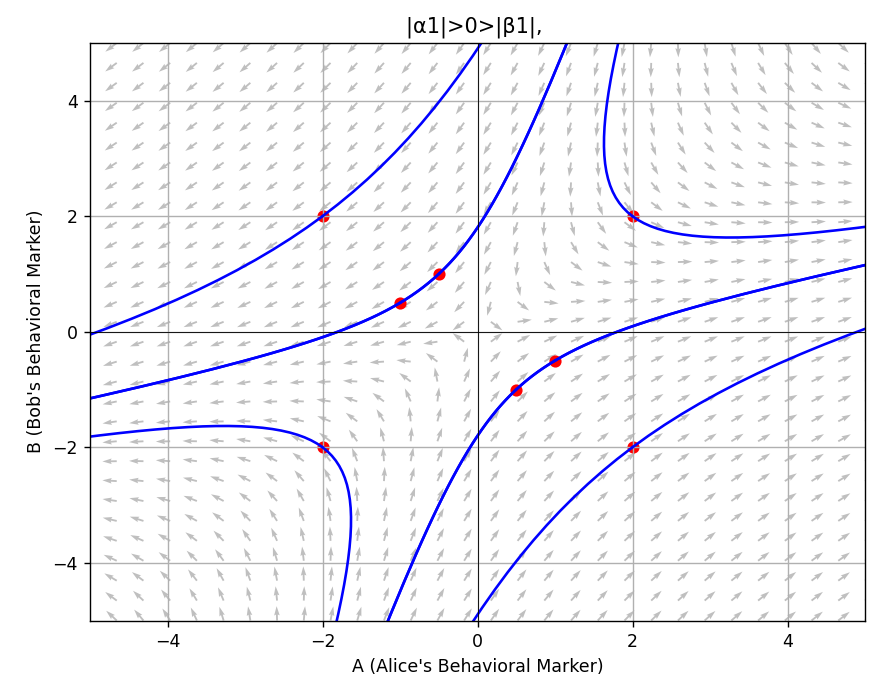} 
        \caption*{(a)}
        \label{fig:ua}
    \end{minipage}
    \hfill
    \begin{minipage}[t]{0.45\textwidth}
        \centering
        \includegraphics[width=\textwidth]{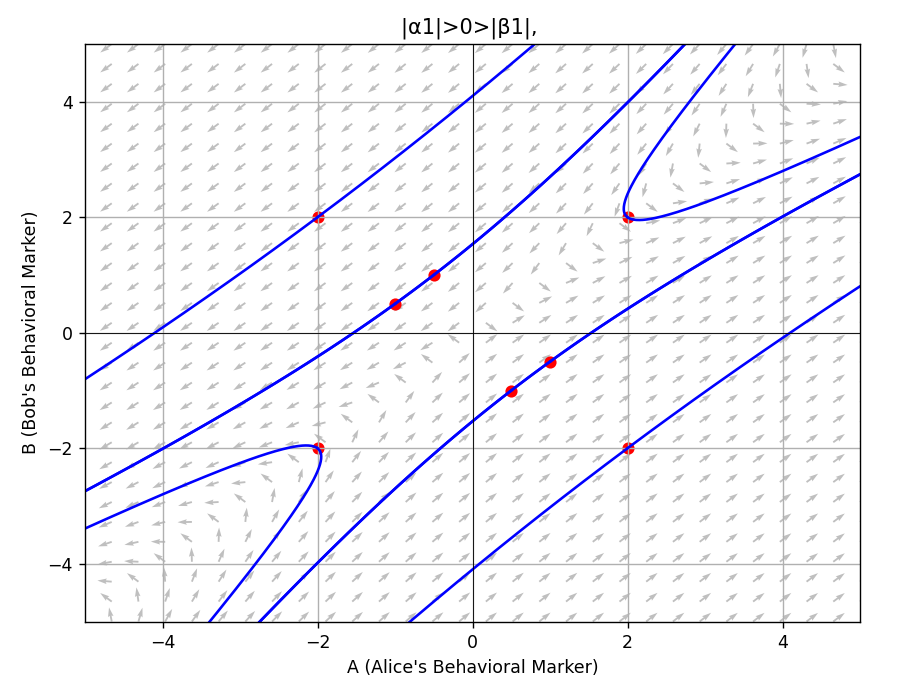} 
        \caption*{(b)}
        \label{fig:eb}
    \end{minipage}
    \caption{Phase portrait of time evolution of behavioral markers for Alice and Bob when (a)$\alpha_{1}=2,\beta_{1}=-1,\alpha_{2}=1,\beta_{2}=-2$ and (b)$\alpha_{1}=1.1,\beta_{1}=-1,\alpha_{2}=1,\beta_{2}=-1.1$ for eight initial conditions of (-2, -2), (2, 2), (-2, 2), (2, -2), (0.5, -1), (-1, 0.5), (1, -0.5), (-0.5, 1) }
    \label{fig:ua_eb}
\end{figure}

 The first and third quadrant demonstrate an important scenario. In the first quadrant both Alice and Bob react positively to each other's behavior and continue doing so, while in the third quadrant both of them react negatively to each other's behaviors and continue doing so.

\item Case 2: Consider a case where $|\alpha_{1}|<|\beta_{1}|$ which implies that $\pm\sqrt{\alpha_1^2 - \alpha_2^2}=\pm i\sqrt{\alpha_2^2 - \alpha_1^2}$.

\item Case 2.1 : Let the first subcase be that $\alpha_{1}>0,\beta_{1}>0$. This means that Alice is a Eager Beaver and Bob is a Hermite. Bob reacts negatively to his own as well as Alice's behavior and Alice reacts positively to hers and Bob's behavior.

\begin{figure}[H]
    \centering
    \begin{minipage}[t]{0.45\textwidth}
        \centering
        \includegraphics[width=\textwidth]{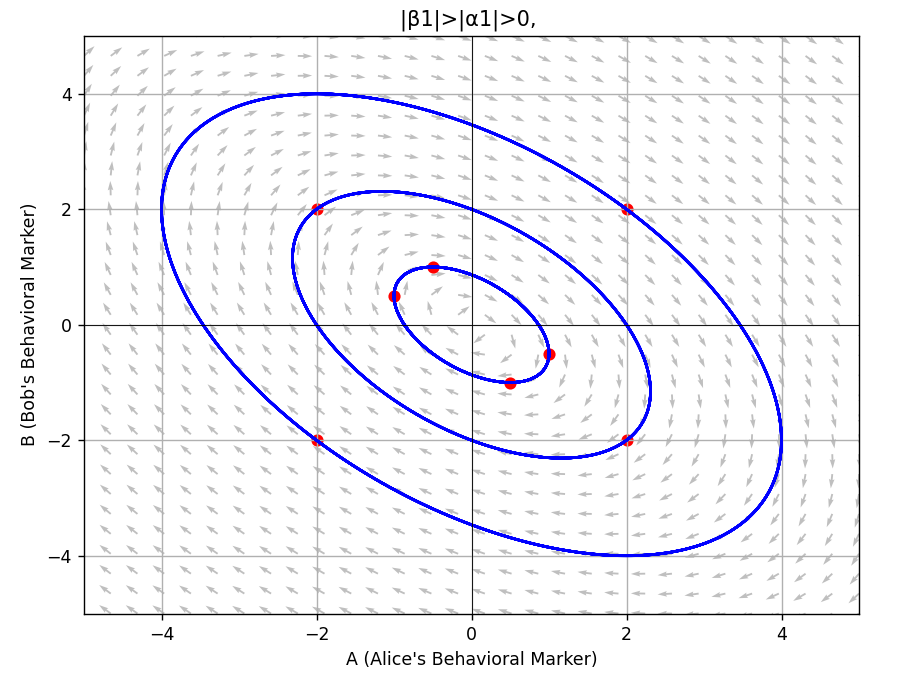} 
        \caption*{(a)}
        \label{fig:ua}
    \end{minipage}
    \hfill
    \begin{minipage}[t]{0.45\textwidth}
        \centering
        \includegraphics[width=\textwidth]{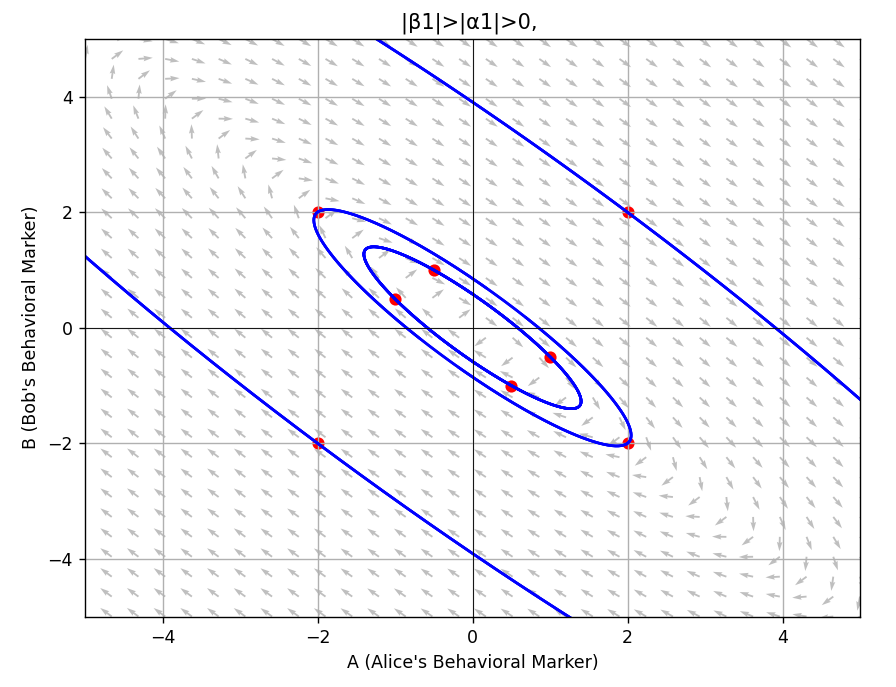} 
        \caption*{(b)}
        \label{fig:eb}
    \end{minipage}
    \caption{Phase portrait of time evolution of behavioral markers for Alice and Bob when (a)$\alpha_{1}=1,\beta_{1}=2,\alpha_{2}=-2,\beta_{2}=-1$ and (b)$\alpha_{1}=1,\beta_{1}=1.1,\alpha_{2}=-1.1,\beta_{2}=-1$ for eight initial conditions of (-2, -2), (2, 2), (-2, 2), (2, -2), (0.5, -1), (-1, 0.5), (1, -0.5), (-0.5, 1) }
    \label{fig:ua_eb}
\end{figure}

\item Case 2.2 : Let the second subcase be that $\alpha_{1}>0,\beta_{1}<0$. This means that Alice is a Narcissistic nerd and Bob is a Cautious lover. Bob reacts negatively to his own behavior and positively to Alice's behavior. Alice reacts positively to her own behavior but negatively to Bob's behavior.

\begin{figure}[H]
    \centering
    \begin{minipage}[t]{0.45\textwidth}
        \centering
        \includegraphics[width=\textwidth]{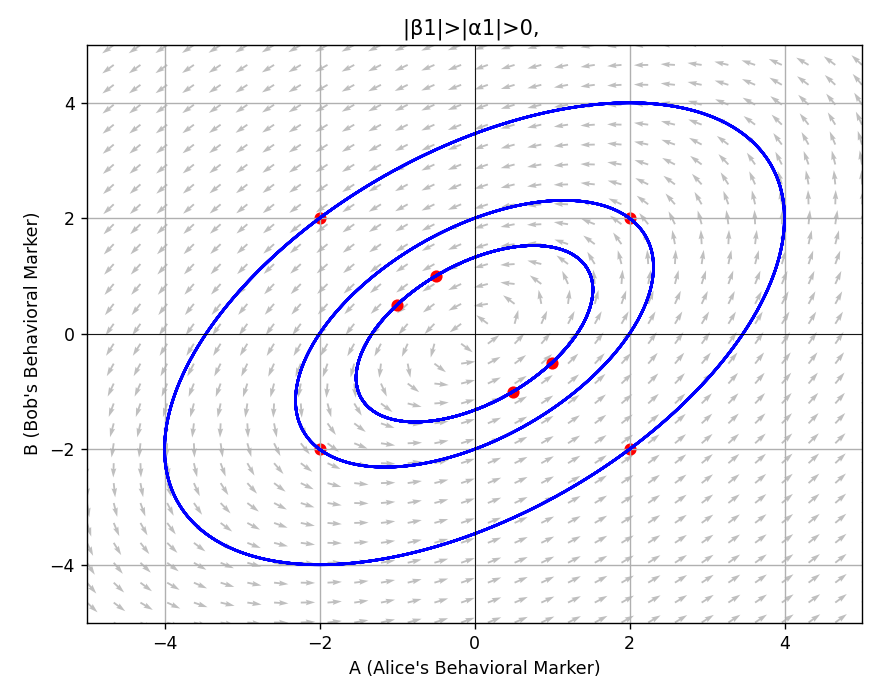} 
        \caption*{(a)}
        \label{fig:ua}
    \end{minipage}
    \hfill
    \begin{minipage}[t]{0.45\textwidth}
        \centering
        \includegraphics[width=\textwidth]{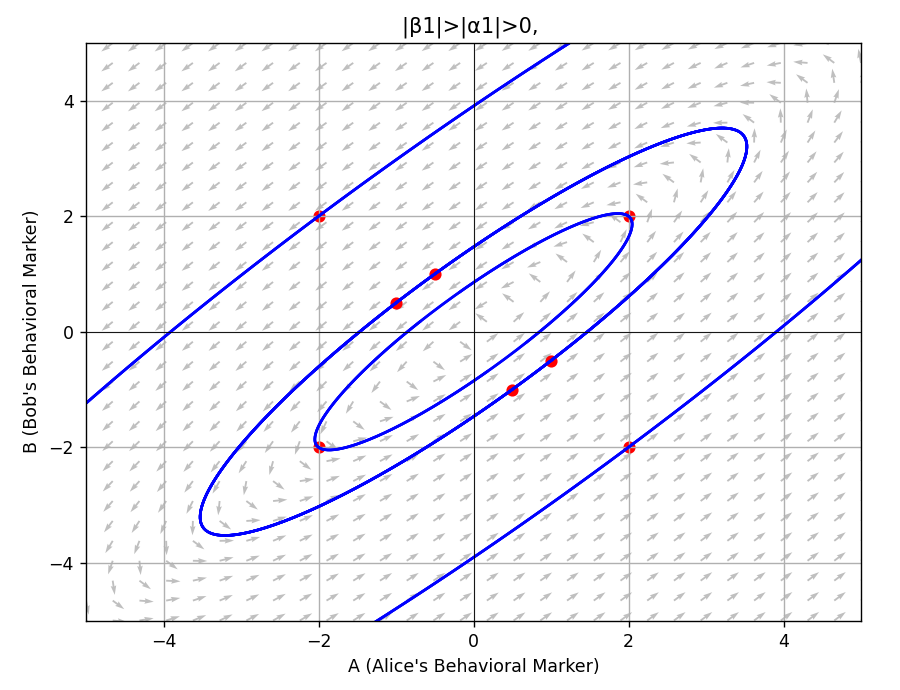} 
        \caption*{(b)}
        \label{fig:eb}
    \end{minipage}
    \caption{Phase portrait of time evolution of behavioral markers for Alice and Bob when (a)$\alpha_{1}=1,\beta_{1}=-2,\alpha_{2}=2,\beta_{2}=-1$ and (b)$\alpha_{1}=1,\beta_{1}=-1.1,\alpha_{2}=1.1,\beta_{2}=-1$ for eight initial conditions of (-2, -2), (2, 2), (-2, 2), (2, -2), (0.5, -1), (-1, 0.5), (1, -0.5), (-0.5, 1). }
    \label{fig:ua_eb}
\end{figure}

For both the above cases, Alice and Bob keep chasing each other being trapped in loops. This leads to a perpetual interaction between Bob and Alice. One important difference in the above two subcases is the motion of the phase field vectors. In the first subcase, this motion is clockwise indicating that as the behavioral markers for Alice and Bob are correlated in quadrants 2,4 and anti-correlated in quadrant 1,3. In the second subcase, the vector field moves anti-clockwise indicating that markers are correlated in quadrants 1,3 and anti-correlated in quadrants 2,4.
Apart from these special cases, where Bob reacts exactly contrary to how Alice reacts, many other general cases can be constructed. 

\begin{figure}[H]
    \centering
    \begin{minipage}[t]{0.45\textwidth}
        \centering
        \includegraphics[width=\textwidth]{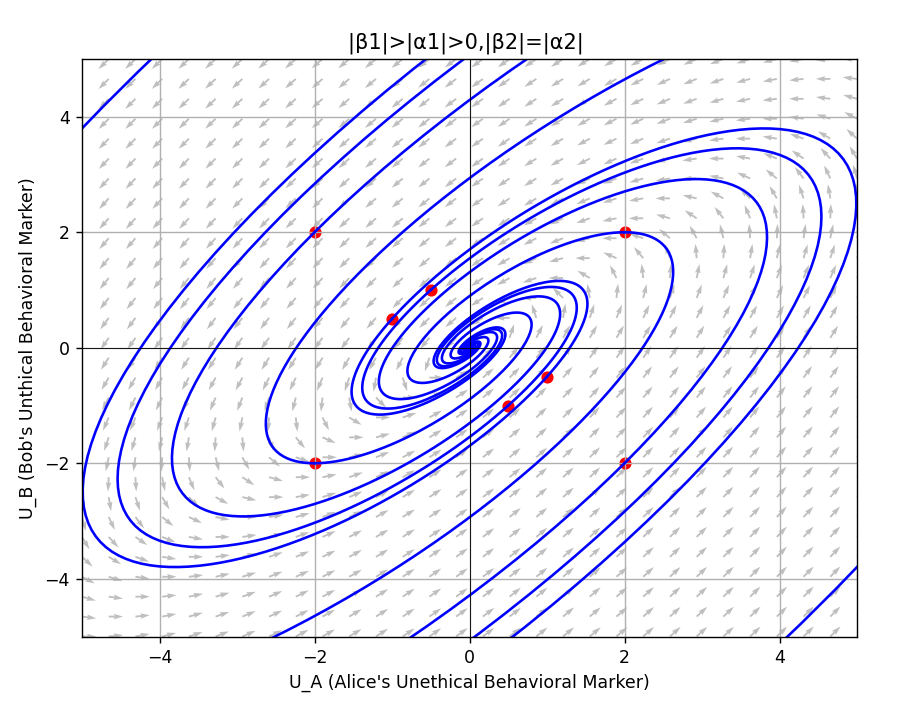} 
        \caption*{(a)}
        \label{fig:ua}
    \end{minipage}
    \hfill
    \begin{minipage}[t]{0.45\textwidth}
        \centering
        \includegraphics[width=\textwidth]{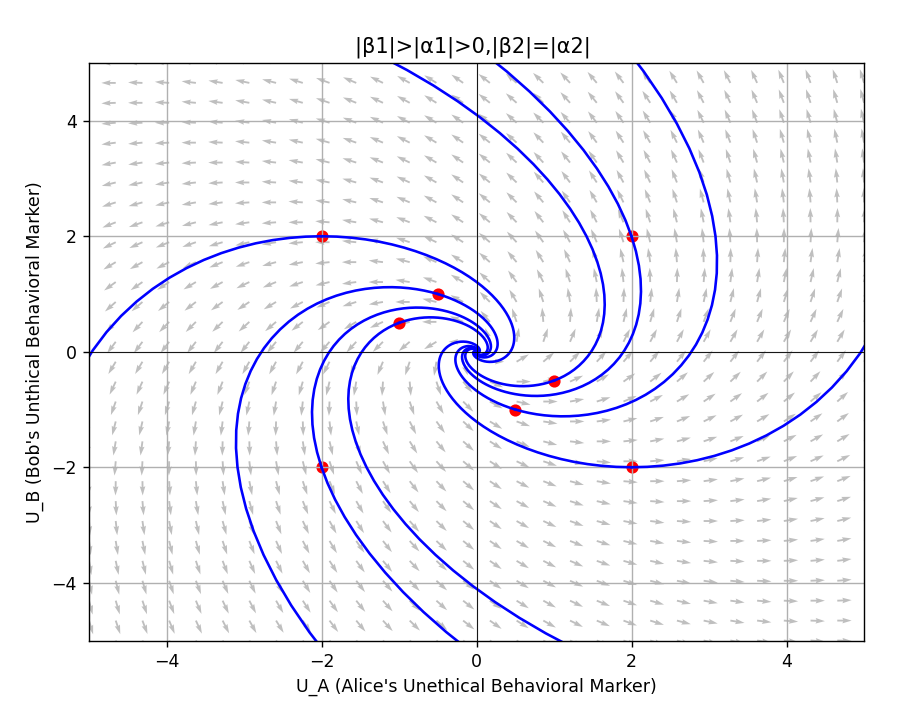} 
        \caption*{(b)}
        \label{fig:eb}
    \end{minipage}
    \caption{Phase portrait of time evolution of behavioral markers for Alice and Bob when (a)$\alpha_{1}=1,\beta_{1}=-2,\alpha_{2}=2,\beta_{2}=-2$ and (b)$\alpha_{1}=1,\beta_{1}=-2,\alpha_{2}=2,\beta_{2}=2$ for eight initial conditions of (-2, -2), (2, 2), (-2, 2), (2, -2), (0.5, -1), (-1, 0.5), (1, -0.5), (-0.5, 1).}
    \label{fig:enter-label}
\end{figure}

\end{itemize}
Sources and sinks can also be generated as is seen from the above example. figure (a) demonstrates action of a sink on trajectories in the phase space. All the trajectories for different initial conditions collapse over time to the final state of $(U_{A},U_{B}) = (0,0) $. On the contrary, figure (b) demonstrates the action of a source where all the trajectories originate from the origin. We refer to these as sink spirals (spiraling in) and source spirals (spiraling out) respectively.

\begin{itemize}

\item 3. The Unethical Duo
\[
\frac{dU_B}{dt} = \alpha_1 U_B + \beta_1 U_A
\]
\[
\frac{dU_A}{dt} = \alpha_2 U_B + \beta_2 U_A
\]
\end{itemize}
The unethical duo case affirms that the unethical behavior of both the players increases indefinitely without the intervention of an external ethical agent. This demands that $\alpha_1,\alpha_2>0$ and $\beta_1,\beta_2>0$ . However if an external ethical agent is introduced as a perturbation to the system, the system may exhibit an interesting behavior.

\begin{figure}[H]
    \centering
    \begin{minipage}[t]{0.45\textwidth}
        \centering
        \includegraphics[width=\textwidth]{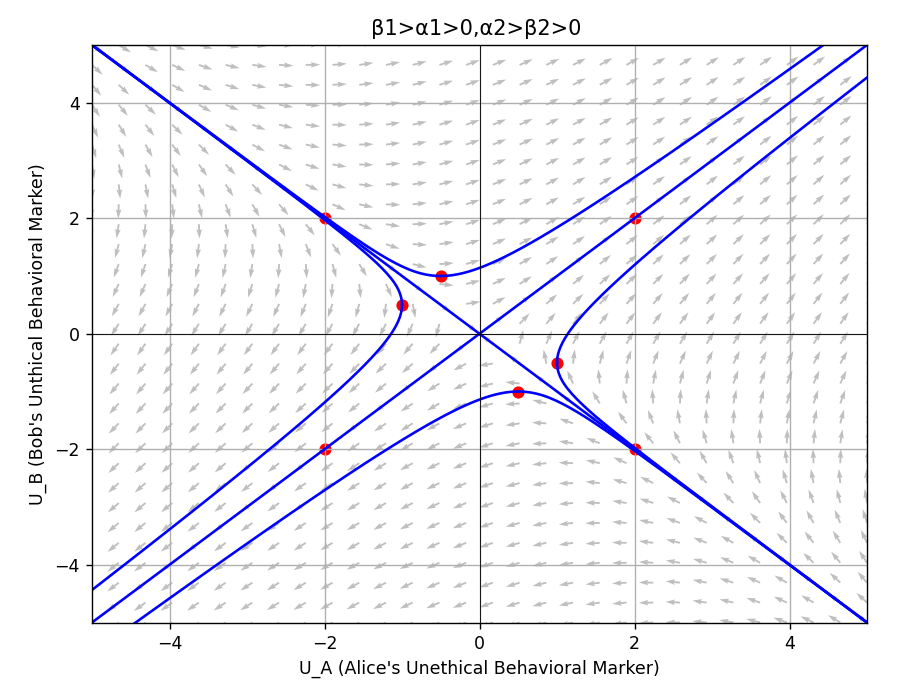} 
        \caption*{(a)}
        \label{fig:ua}
    \end{minipage}
    \hfill
    \begin{minipage}[t]{0.45\textwidth}
        \centering
        \includegraphics[width=\textwidth]{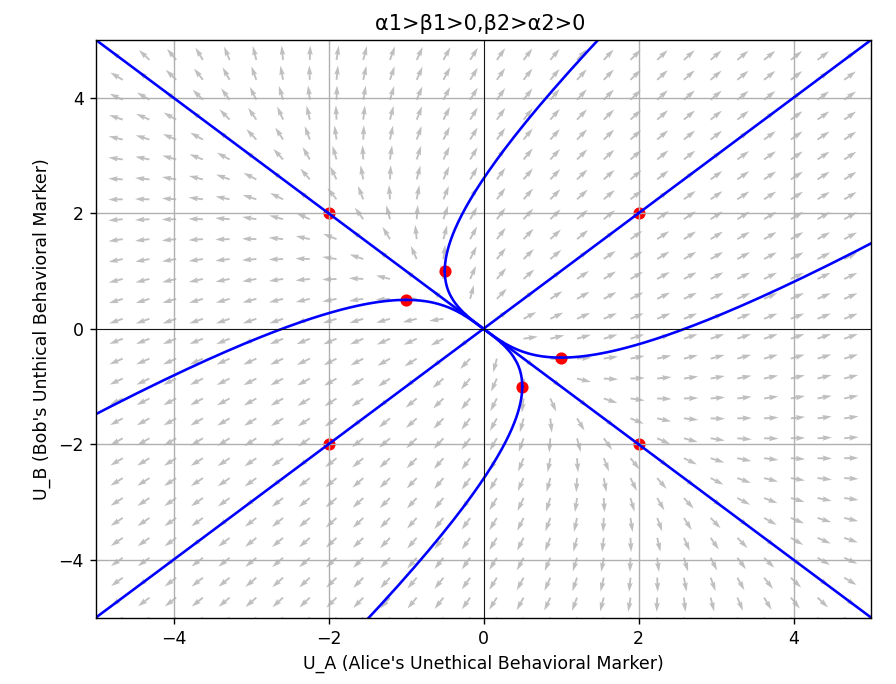} 
        \caption*{(b)}
        \label{fig:eb}
    \end{minipage}
    \caption{The figure represents the phase portrait of the unethical duo for (a) $\alpha_{1}=\beta_{2}=1$,$\beta_{1}=\alpha_{2}=2$ and (b)$\alpha_{1}=\beta_{2}=2$,$\beta_{1}=\alpha_{2}=1$ for eight initial conditions of (-2, -2), (2, 2), (-2, 2), (2, -2), (0.5, -1), (-1, 0.5), (1, -0.5), (-0.5, 1)}
    \label{fig:ua_eb}
\end{figure}

The first scenario (a) here shows dynamics of the relation when the cross coupling is stronger than self coupling. This means that each of Alice and Bob are influenced more by the unethical behavior of their counterpart than their own behavior. In $2^{nd}$ and $3^{rd}$ quadrants, trajectories are moving towards the origin and in the remaining two quadrants they are moving away from origin. In a dynamical system, the nature of equilibrium points is determined by the eigenvalues of the Jacobian matrix \( J \). The determinant of \( J \), given by \( \det(J) \), represents the product of its eigenvalues. When \( \det(J) < 0 \), the eigenvalues have opposite signs, indicating that the equilibrium point has one stable and one unstable direction. This results in a saddle point, where trajectories approach along the stable eigenvector and diverge along the unstable eigenvector. In contrast, for \( \det(J) > 0 \), the equilibrium could be either a node or a spiral, depending on the trace of \( J \). We have already demonstrated nodes and sink/source trajectories.

The Jacobian matrix \( J \) for the above case is given by:
\[
J = \begin{bmatrix} 
2 & 1 \\ 
2 & 1 
\end{bmatrix}
\]

For the above Jacobian, $Tr(J) =3$ and $det(J)=0$. This indicates an unstable equilibrium and degeneracies in solutions. Same is true for the second scenario (b) also. To obtain unique solutions and a clear saddle point,$det(J)<0$ is a necessary condition. To mimic this situation, we consider two more cases.

\begin{figure}[H]
    \centering
    \begin{minipage}[t]{0.45\textwidth}
        \centering
        \includegraphics[width=\textwidth]{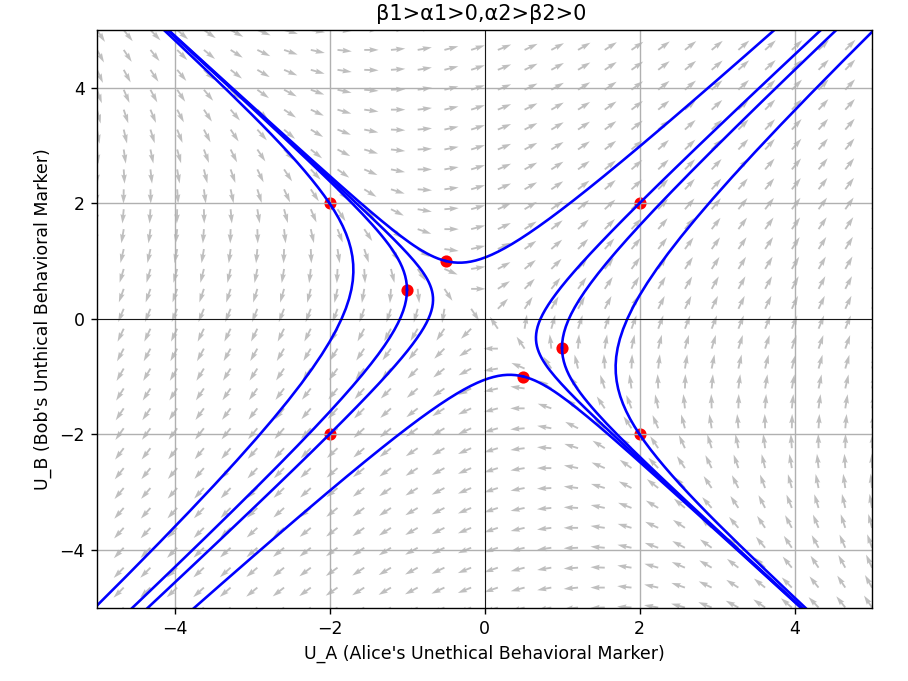} 
        \caption*{(a)}
        \label{fig:ua}
    \end{minipage}
    \hfill
    \begin{minipage}[t]{0.45\textwidth}
        \centering
        \includegraphics[width=\textwidth]{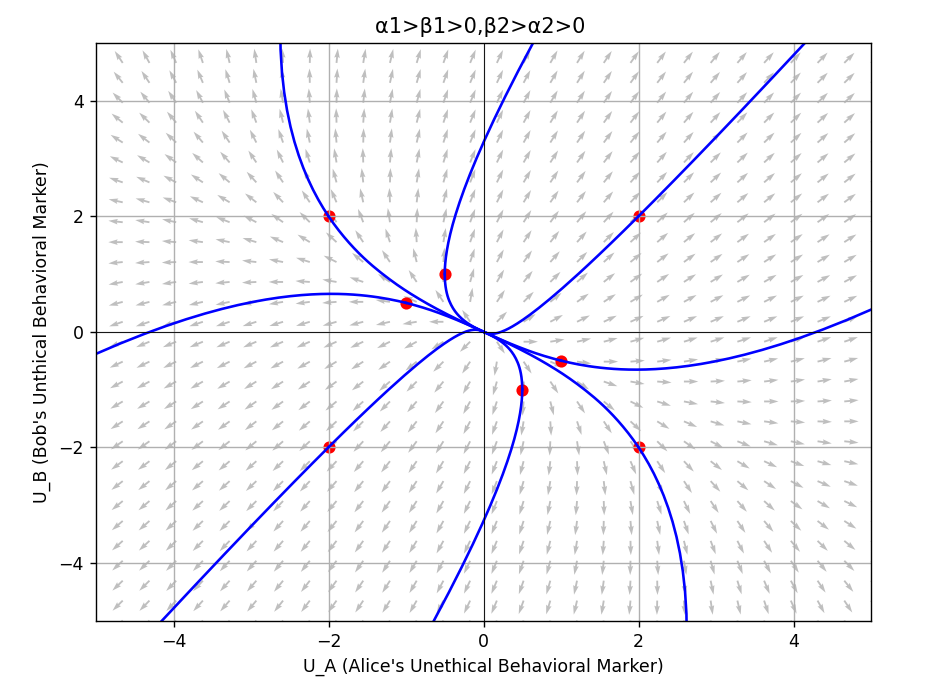} 
        \caption*{(b)}
        \label{fig:eb}
    \end{minipage}
    \caption{The figure represents the phase portrait of the unethical duo for (a) $\alpha_{1}=1,\beta_{1}=2$,$\beta_{2}=1,\alpha_{2}=3,$$Tr(J)=2$,$det(J)=-5$ and (b)$\alpha_{1}=2,\beta_{1}=1$,$\beta_{2}=3,\alpha_{2}=1,Tr(J)=5, det(J)=5$, for eight initial conditions of (-2, -2), (2, 2), (-2, 2), (2, -2), (0.5, -1), (-1, 0.5), (1, -0.5), (-0.5, 1). For scenario (a).}
    \label{fig:ua_eb}
\end{figure}

The above system of equations does not take into account any external perturbation. Consider that another ethical player Carl (C) comes into picture. $E_C$ now acts as an ethical perturbation for unethical Alice ($U_A$) and unethical Bob ($U_B$). For the sake of simplicity we consider that Alice and Bob do not affect Carl in any way but Carl can act as a perturbation for both Alice and Bob. Given the scenario, we have three coupled differential equation describing this dynamics. We analyze this system case wise.

\begin{itemize}

\item Case 1: Ethical Carl acts as a perturbation only for Alice and Carl is unaffected by behavior of Alice and Bob. This system can be effectively modeled by the following three differential equations;

\[
\frac{dU_B}{dt} = \alpha_1 U_B + \beta_1 U_A \]
\[
\frac{dU_A}{dt} = \alpha_2 U_B + \beta_2 U_A+\gamma_2 E_C
\]
\[
\frac{dE_C}{dt} = \gamma_3E_C
\]

\end{itemize}

We solve the given system of first-order differential equations using the Euler method, we discretize time with a small step size \( \Delta l \) and approximate the derivatives using forward differences. The system
can be solved iteratively by updating each variable at discrete time steps \( l_n = l_0 + n\Delta l \) using the Euler update rule.On analytically computing this system of equations, we derive phase portrait trajectories for Alice, Bob and Carl. This setting is a three player game with unidirectional or bidirectional effect of the third player as a perturbation entity. This setting mimics a real life time evolution of organizational behavior with individuals of conflicting interest.An interesting dynamics can emerge out of Carl. For example, if $\gamma_{3}<0$, Carl has a decaying ethical behavior and Carl's ethical perturbation may die out soon. However, $\gamma_{3}>0$ will keep the perturbation evolve over longer time periods. We demonstrate a source/sink like 3D trajectories when Carl has a dying out ethical behavior and Alice is a Hermite. 

\begin{figure}[H]
    \centering
    \begin{minipage}[t]{0.45\textwidth}
        \centering
        \includegraphics[width=\textwidth]{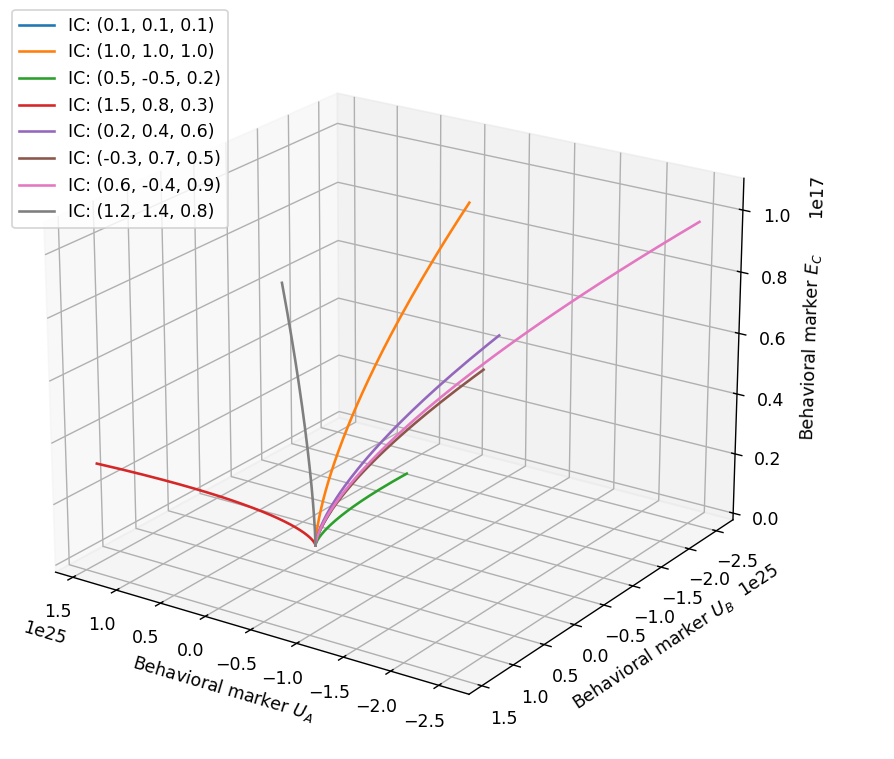} 
        \caption*{(a)}
        \label{fig:ua}
    \end{minipage}
    \hfill
    \begin{minipage}[t]{0.45\textwidth}
        \centering
        \includegraphics[width=\textwidth]{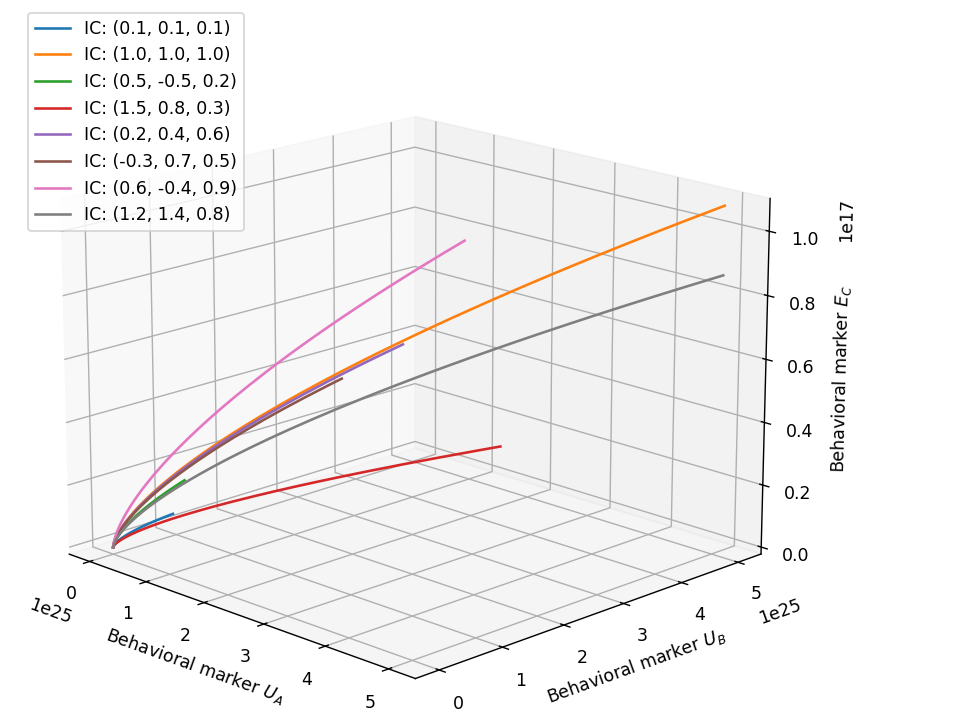} 
        \caption*{(b)}
        \label{fig:eb}
    \end{minipage}
    \hfill
    \begin{minipage}[t]{0.45\textwidth}
        \centering
        \includegraphics[width=\textwidth]{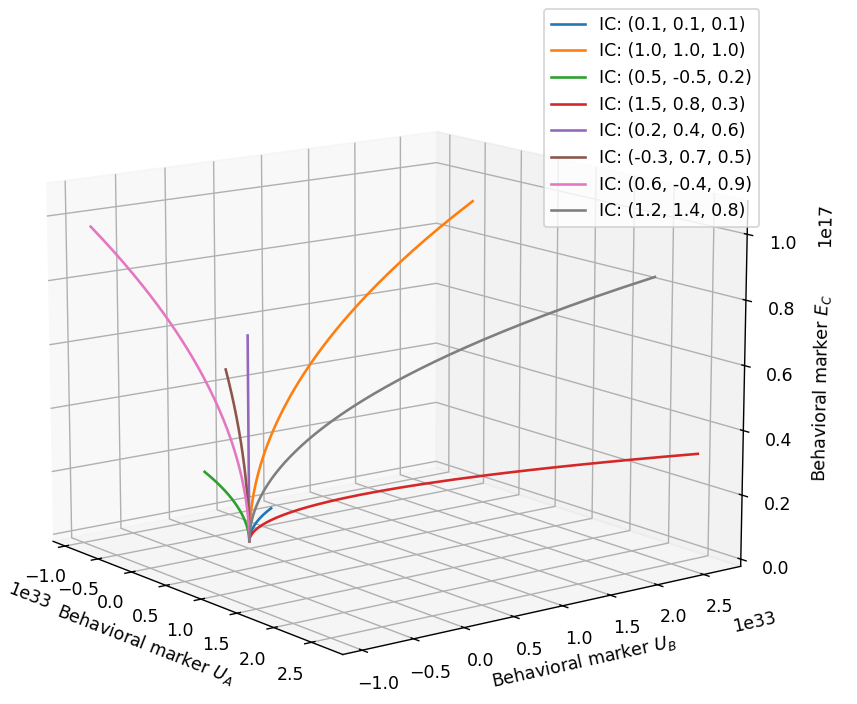} 
        \caption*{(c)}
        \label{fig:eb}
    \end{minipage}
    \hspace{12mm}
    \begin{minipage}[t]{0.45\textwidth}
        \centering
        \includegraphics[width=\textwidth]{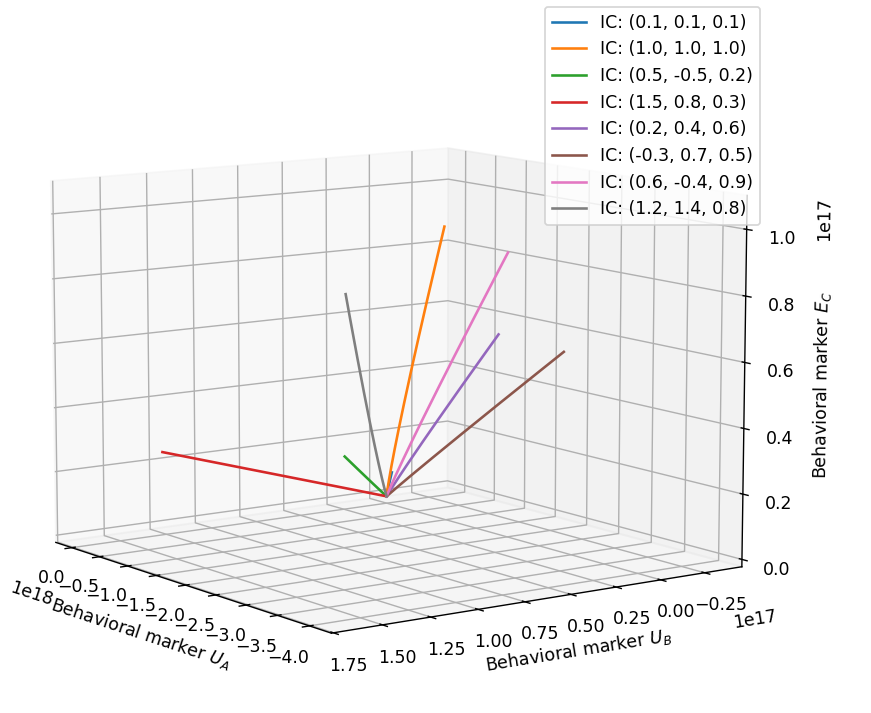} 
        \caption*{(d)}
        \label{fig:eb}
    \end{minipage}
    \hfill
    \begin{minipage}[t]{0.45\textwidth}
        \centering
        \includegraphics[width=\textwidth]{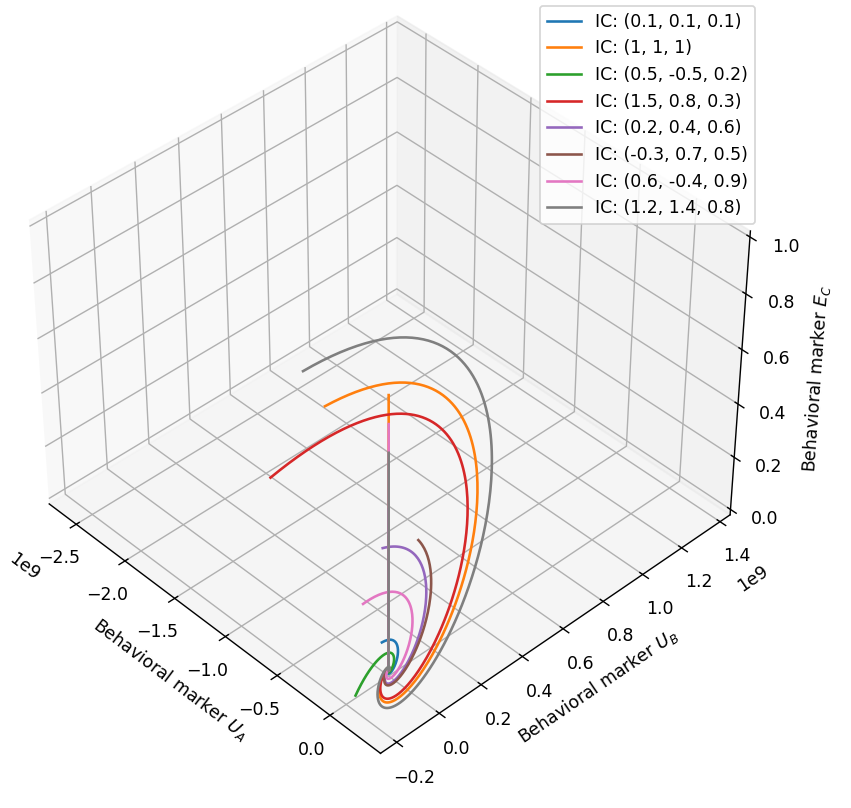} 
        \caption*{(e)}
        \label{fig:eb}
    \end{minipage}
    \hfill
    \caption{\footnotesize{The figure represents the phase portrait of the unethical duo with ethical perturbation of carl for (a) $\alpha_{1}=1,\beta_{1}=2$,$\beta_{2}=1,\alpha_{2}=2,\gamma_{2}=-3,\gamma_{3}=2$, (b)$\alpha_{1}=1,\beta_{1}=2$,$\beta_{2}=2,\alpha_{2}=1,\gamma_{2}=3,\gamma_{3}=2$,(c)$\alpha_{1}=2,\beta_{1}=2$,$\beta_{2}=2,\alpha_{2}=2,\gamma_{2}=-2,\gamma_{3}=2$ and (d)$\alpha_{1}=2,\beta_{1}=0$,$\beta_{2}=0,\alpha_{2}=2,\gamma_{2}=-2,\gamma_{3}=2$, (e)$\alpha_{1}=3,\beta_{1}=2$,$\beta_{2}=-1,\alpha_{2}=-3,\gamma_{2}=2,\gamma_{3}=-2$ for eight initial conditions of (0.1, 0.1, 0.1), (1.0, 1.0, 1.0), (0.5, -0.5, 0.2), (1.5, 0.8, 0.3), (0.2, 0.4, 0.6), (-0.3, 0.7, 0.5), (0.6, -0.4, 0.9), (1.2, 1.4, 0.8).}}
    \label{fig:ua_eb}
\end{figure}

In all the above scenarios, there is a singular influence of carl only on Alice. This generates a dynamics where the ethical behavior of Carl only affect Alice and Bob is entirely unaffected by this behavior.
\begin{itemize}
\item Case 2:Considering that Carl also affects Bob, the new dynamics of the system is given by the following set of differential equations;
\end{itemize}
\[
\frac{dU_B}{dt} = \alpha_1 U_B + \beta_1 U_A +\gamma_{1} E_C\]
\[
\frac{dU_A}{dt} = \alpha_2 U_B + \beta_2 U_A+\gamma_2 E_C
\]
\[
\frac{dE_C}{dt} = \gamma_3E_C
\]

\begin{figure}[H]
    \centering
    \begin{minipage}[t]{0.45\textwidth}
        \centering
        \includegraphics[width=\textwidth]{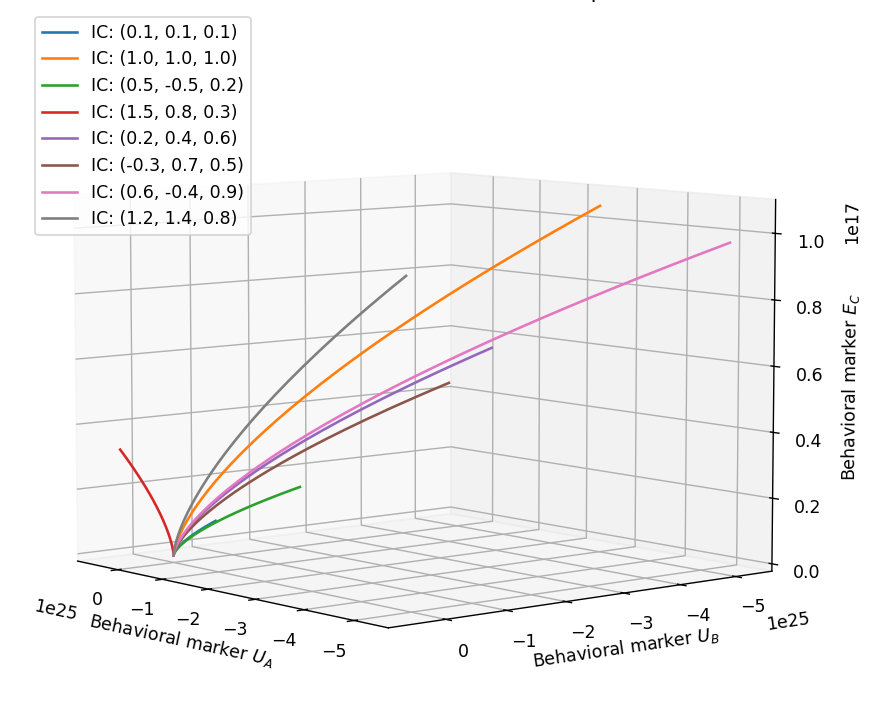} 
        \caption*{(a)}
        \label{fig:ua}
    \end{minipage}
    \hfill
    \begin{minipage}[t]{0.45\textwidth}
        \centering
        \includegraphics[width=\textwidth]{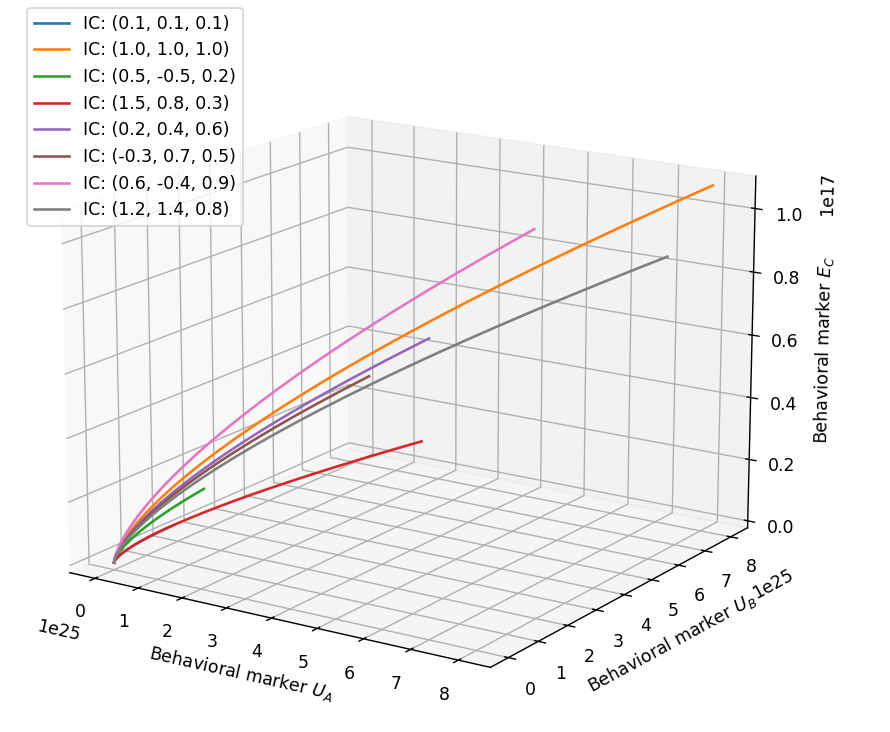} 
        \caption*{(b)}
        \label{fig:eb}
    \end{minipage}
    \hfill
    \begin{minipage}[t]{0.45\textwidth}
        \centering
        \includegraphics[width=\textwidth]{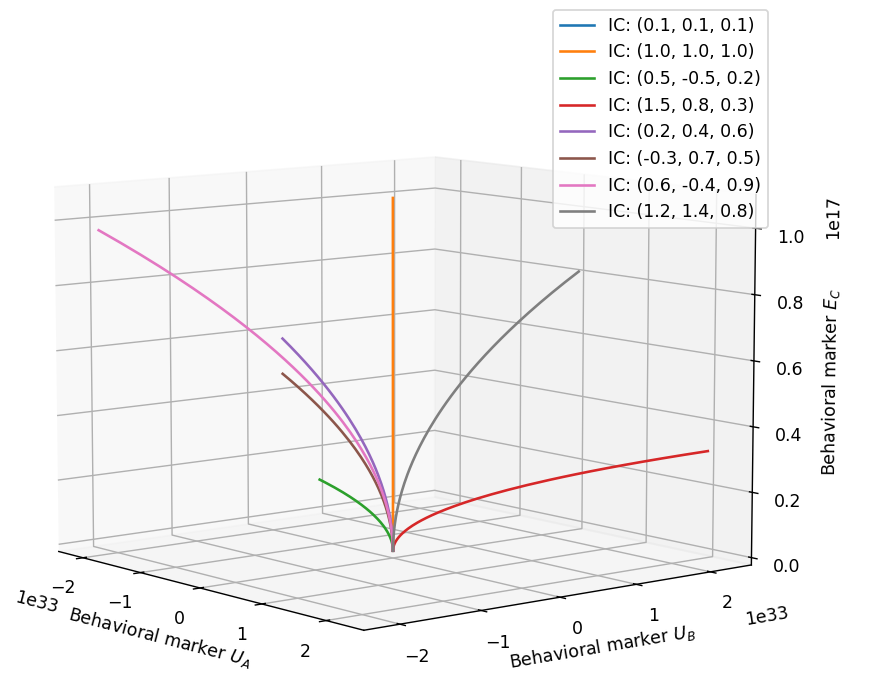} 
        \caption*{(c)}
        \label{fig:eb}
    \end{minipage}
    \hspace{12mm}
    \begin{minipage}[t]{0.45\textwidth}
        \centering
        \includegraphics[width=\textwidth]{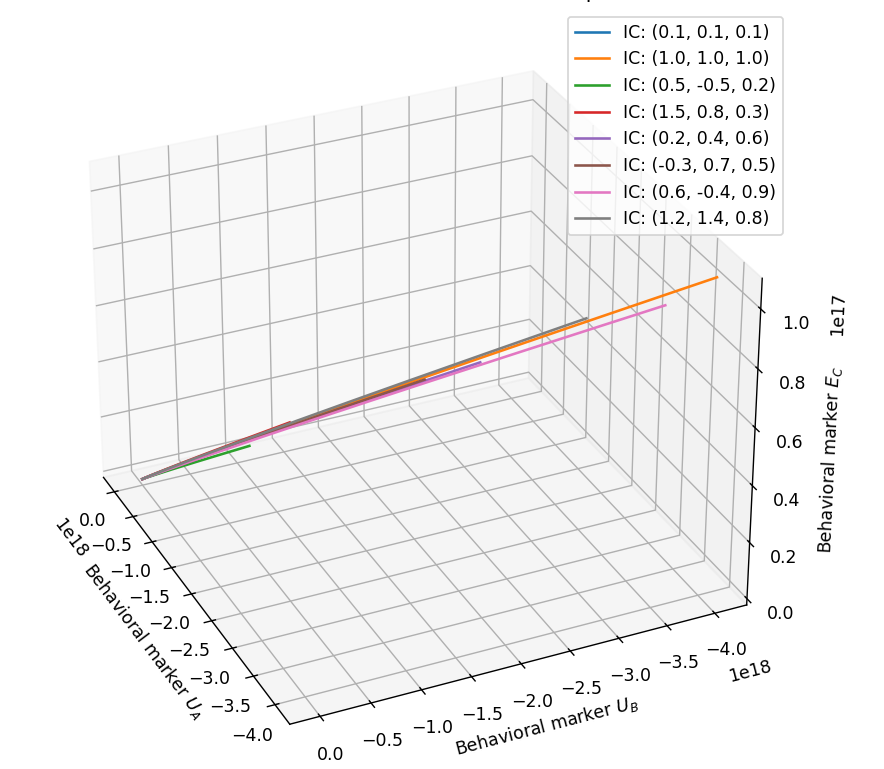} 
        \caption*{(d)}
        \label{fig:eb}
    \end{minipage}
    \hfill
    \caption{The figure represents the phase portrait of the unethical duo with ethical perturbation of carl for (a) $\alpha_{1}=1,\beta_{1}=2$,$\beta_{2}=1,\alpha_{2}=2,\gamma_{1}=-3,\gamma_{2}=-3,\gamma_{3}=2$, (b)$\alpha_{1}=1,\beta_{1}=2$,$\beta_{2}=2,\alpha_{2}=1,\gamma_{1}=3,\gamma_{2}=3,\gamma_{3}=2$,(c)$\alpha_{1}=2,\beta_{1}=2$,$\beta_{2}=2,\alpha_{2}=2,\gamma_{1}=-2,\gamma_{2}=-2,\gamma_{3}=2$ and (d)$\alpha_{1}=2,\beta_{1}=0$,$\beta_{2}=0,\alpha_{2}=2,\gamma_{1}=-2,\gamma_{2}=-2,\gamma_{3}=2$, for eight initial conditions of (0.1, 0.1, 0.1), (1.0, 1.0, 1.0), (0.5, -0.5, 0.2), (1.5, 0.8, 0.3), (0.2, 0.4, 0.6), (-0.3, 0.7, 0.5), (0.6, -0.4, 0.9), (1.2, 1.4, 0.8).}
    \label{fig:ua_eb}
\end{figure}

A profound effect can be seen for (d) scenario with singular effect of Carl only on Alice and effect of carl on Alice as well as Bob. A special case of above system of differential equations can be analyzed by letting $\gamma_{3}=0$. This suggests that the ethical behavior of Carl does not change over time and is constant. For such case, phase trajectories are expected to be nullclines at z coordinates of each initial condition. This can be easily verified.

\textbullet{ 4. The Ethical Duo}

\[
\frac{dE_B}{dt} = \alpha_1 E_B + \beta_1 E_A
\]

\[
\frac{dE_A}{dt} = \alpha_2 E_A + \beta_2 E_B
\]
Similar to the unethical duo case, this case affirms that the ethical behavior of both players increases indefinitely without the intervention of an external unethical agent. This case is also parametrized by $\alpha_1,\alpha_2>0$ and $\beta_1,\beta_2>0$ .

This scenario can be analyzed in exactly the same manner as the ethical duo scenario. An unethical perturbation in this case will generate exactly same effect as the ethical perturbation in the above scenario.

\section{Conclusion}

In this paper, we have demonstrated two efficient techniques to study the time evolution of behavioral markers of individuals who are a part of organization. The matrix formulation based polytope approach can be efficiently utilized to study relatively simple type of evolution where the ethical and unethical behaviors are constrained by one or a set of inequalities. Different polytopes can be generated for different games characterized by nature of inequalities. Each of these polytopes encompass all possible solutions of that game with the inequalities being faucets of that polytope. The differential approach can be extended to a N player game and can be used to generate N dimensional phase portrait of the game. We have also demonstrated the effect of a perturbation in a two player game,leading to interesting source and sink trajectories which can be modeled alternatively as effects of external parameters. Differential approach is preferred to study time of complicated multiplayer games with randomized or tailored interdependence which are modeled by unique parameter set of each player with maximum cardinality of each set being N for a N player game. Through this work we have established a robust mathematical treatment of problems of ethics and alike.

\section*{References}

\begin{enumerate}
    \item Bommer, M., Gratto, C., Gravander, J. et al. A behavioral model of ethical and unethical decision making. \textit{J. Bus. Ethics} \textbf{6}, 265–280 (1987). \href{https://doi.org/10.1007/BF00382936}{https://doi.org/10.1007/BF00382936}
    
    \item Biggar, O., Shames, I. The graph structure of two-player games. \textit{Sci Rep} \textbf{13}, 1833 (2023). 
    
    \href{https://doi.org/10.1038/s41598-023-28627-8}{https://doi.org/10.1038/s41598-023-28627-8}
    
    \item Von Neumann, J., Morgenstern, O. \textit{Theory of Games and Economic Behavior} (Princeton University Press, 1944).
    
    \item Nash, J. Non-cooperative games. \textit{Ann. Math.} \textbf{20}, 286–295 (1951).
    
    \item Cruz, J.B., Simaan, M.A. Ordinal Games and Generalized Nash and Stackelberg Solutions. \textit{J. Optim. Theory Appl.} \textbf{107}, 205–222 (2000). \href{https://doi.org/10.1023/A:1026476425031}{https://doi.org/10.1023/A:1026476425031}
    
    \item Strogatz, S. \textit{Nonlinear Dynamics and Chaos} (Westview Press, 2015).
    
    \item Gou, Z., Li, Y. Prisoner’s dilemma game model Based on historical strategy information. \textit{Sci Rep} \textbf{13}, 1 (2023). \href{https://doi.org/10.1038/s41598-022-26890-9}{https://doi.org/10.1038/s41598-022-26890-9}
    
    \item Heuer, L. and Orland, A. Cooperation in the Prisoner’s Dilemma: an experimental comparison between pure and mixed strategies. \textit{R. Soc. Open Sci.} \textbf{6}, 182142 (2019). \href{http://doi.org/10.1098/rsos.182142}{http://doi.org/10.1098/rsos.182142}
    
    \item Montero-Porras, E., Grujić, J., Fernández Domingos, E. et al. Inferring strategies from observations in long iterated Prisoner’s dilemma experiments. \textit{Sci Rep} \textbf{12}, 7589 (2022). \href{https://doi.org/10.1038/s41598-022-11654-2}{https://doi.org/10.1038/s41598-022-11654-2}
    
    \item Roth, A.E., Murnighan, J.K. Equilibrium behavior and repeated play of the prisoner’s dilemma. \textit{J. Math. Psych.} \textbf{17}, 189-198 (1978). \href{https://doi.org/10.1016/0022-2496(78)90030-5}{https://doi.org/10.1016/0022-2496(78)90030-5}
    
    \item Holt, C.A., Roth, A.E. The Nash equilibrium: A perspective. \textit{Proc. Natl. Acad. Sci. U.S.A.} \textbf{101}(12), 3999-4002 (2004). \href{https://doi.org/10.1073/pnas.0308738101}{https://doi.org/10.1073/pnas.0308738101}
    
    \item Strogatz, S.H. Love affairs and differential equations. \textit{Mathematics Magazine} \textbf{61}(1), 35 (1988).

    \item Simaan, M., and Cruz, J. B., JR. On the Stackelberg Strategy in Nonzero-Sum Games. \textit{J. Optim. Theory Appl.} \textbf{11}, 533–555 (1973). \href{https://doi.org/10.1007/BF00935665}{https://doi.org/10.1007/BF00935665}

\end{enumerate}

\end{document}